\documentclass[%
 aapm,
 mph,%
 amsmath,amssymb,
 reprint,
 nofootinbib,
]{revtex4-2}

\usepackage{graphicx}
\usepackage{dcolumn}
\usepackage{bm}
\usepackage{textcomp}%
\usepackage{xcolor}
\usepackage[mathlines]{lineno}
\modulolinenumbers[5]

\begin{document}

\preprint{AAPM/123-QED}

\title[Constraining Axion Mass ]{Constraining Axion Mass through Gamma-ray Observations of Pulsars}

\author{Sheridan J. Lloyd}
\email{sheridan.j.lloyd@durham.ac.uk}
\author{Paula M. Chadwick}%
\email{p.m.chadwick@durham.ac.uk}
\author{Anthony M. Brown}%
\email{anthony.brown@durham.ac.uk}

\affiliation{ 
Centre for Advanced Instrumentation, Dept. of Physics, University of Durham, South Road, Durham, DH1 3LE, UK
}%

\date{\today}

\begin{abstract}
We analyze 9 years of \textsc{pass} 8 \textit{Fermi}-LAT data in the 60$-$500 MeV range and determine flux upper limits (UL) for 17 gamma-ray dark pulsars as a probe of axions produced by nucleon-nucleon Bremsstrahlung in the pulsar core. Using a previously published axion decay gamma-ray photon flux model for pulsars  \textcolor{black}{which relies on a high} core temperature of 20 MeV, we improve the determination of the UL axion mass (\textit{m\textsubscript{a}}), at 95 percent confidence level, to 9.6 $\times$  10\textsuperscript{-3} eV, which is a factor of 8 improvement on previous results. We show that the axion emissivity (energy loss rate per volume) at realistic lower pulsar core temperatures of 4 MeV or less \textcolor{black}{is reduced to such an extent that axion emissivity and the gamma-ray signal becomes negligible}.  We consider an alternative emission model based on energy loss rate per mass \textcolor{black}{to allow \textit{m\textsubscript{a}} to be constrained with \textit{Fermi}-LAT observations}. This model yields a plausible UL \textit{m\textsubscript{a}} of 10\textsuperscript{-6} eV for pulsar core temperature \textless 0.1 MeV but \textcolor{black}{knowledge of the} extent of axion to photon conversion in the pulsar \textit{B} field would be required to make a precise UL axion mass determination. The peak of axion flux is likely to produce gamma-rays in the  $\leq$ 1 MeV energy range and so future observations with medium energy gamma-ray missions, such as AMEGO and e-ASTROGAM, will be vital to further constrain UL \textit{m\textsubscript{a}}.
\end{abstract}

\maketitle

\section{\label{sec:level1}Introduction\protect\\}

The axion, a Nambu-Goldstone boson, is a solution to the strong CP problem of QCD and a plausible cold dark matter candidate [\onlinecite{PhysRevLett.38.1440,PhysRevLett.40.223,Dine:1981rt}]. The mass of the axion \textit{m\textsubscript{a}} can be constrained by astrophysical arguments such as the duration of the neutrino burst of SN-1987A (\textit{m\textsubscript{a}} \textless5 $\times$ 10\textsuperscript{-3} eV) [\onlinecite{PhysRevD.56.2419}] or by direct detection experiments such as ADMX [\onlinecite{RYBKA201414}] where Galactic halo axions convert to microwave photons in a magnetic field, excluding \textit{m\textsubscript{a}} in the range (1.9-3.53) $\times$ 10\textsuperscript{-6} eV [\onlinecite{PhysRevLett.104.041301,PhysRevD.74.012006,PhysRevLett.80.2043,PhysRevD.84.121302,ADMX}]. The authors of [\onlinecite {PhysRevD.93.065044}] have used cooling simulations, combined with surface temperature measurements of 4 thermal X-ray emitting pulsars (PSRs), to determine \textit{m\textsubscript{a}} \textless(0.06-0.12 eV). In the gamma-ray regime, the authors of [\onlinecite{RN131}] have used 5 years of \textsc{pass} 7 \textit{Fermi}-LAT gamma-ray observations of radiative axion decay in 4 nearby PSRs to constrain \textit{m\textsubscript{a}} \textless0.079 eV.

The latest data release of the \textit{Fermi}-LAT is now \textsc{pass} 8, which incorporates improvements to further reduce gamma-ray background uncertainty, improve instrument effective area and point spread function (PSF) and to permit low-energy analysis down to 60 MeV. In this paper we will seek to refine the work of [\onlinecite{RN131}] to take advantage of the improved low-energy analysis in \textsc{pass} 8,  coupled with improved photon statistics (9 years of event data) and a larger sample of 17 gamma-ray dark PSRs.  This should allow a more robust determination of UL \textit{m\textsubscript{a}} than was possible previously. 

This paper is structured as follows. In Section \ref{sec:Phenomenology} we describe the phenomenology of the axion and its production in neutron stars. In Section \ref{sec:GCSelection} we describe the criteria used to select pulsars for analysis. In Section \ref{sec:Analysis} we describe our analysis method for the determination of gamma-ray upper limits from the pulsar sample. In Section \ref{sec:Results} we present UL energy and photon flux determinations for the pulsar sample and from these  derive the axion mass upper limit \textit{m\textsubscript{a}} by two independent methods. In Section \ref{sec:Discussion} we discuss the validity of the UL \textit{m\textsubscript{a}} determination with respect to pulsar core temperature. Finally in Section \ref{sec:Conclusion} we summarise our findings and make suggestions for future work.

\section{Phenomenology}
\label{sec:Phenomenology}

In this section we discuss the mechanism for axion production in degenerate pulsar cores and describe how this process is modelled through a spin structure function. We then restate how the axion emissivity or energy loss rate per volume is expressed in terms of this spin structure function. We use a published astrophysical model for the photon flux arising from axion emission and decay in pulsars to derive an expression for UL axion mass. Finally we derive an alternative expression for UL axion mass by using the expected energy loss rate per mass due to axion production to give an expected gamma-ray luminosity for a canonical pulsar and then equate this to the measured gamma-ray upper limits of the pulsars we consider.

\begin{figure}
	\includegraphics[width=\columnwidth]{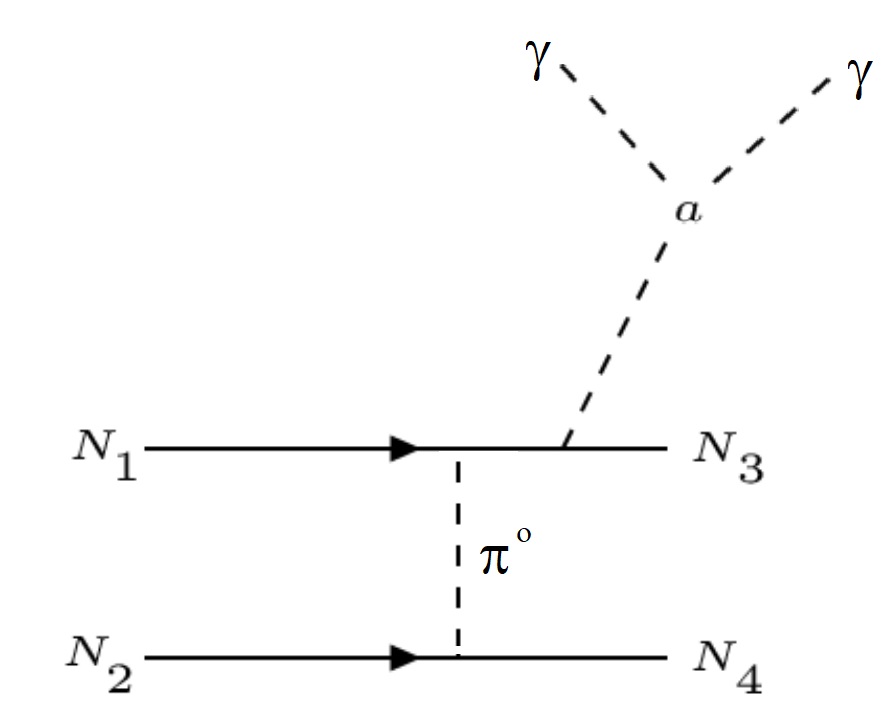}
    \caption{Feynman diagram depicting the nucleon-nucleon Bremsstrahlung process which produces axions. Incoming nucleons N\textsubscript{1},\textsubscript{2} undergo a one-pion exchange producing an axion \textit{a} and outgoing nucleons N\textsubscript{3},\textsubscript{4} with  different energy and momenta from those of N\textsubscript{1},\textsubscript{2}. The axion undergoes radiative (conservative) decay to two gamma-ray photons. }
    
    \label{fig:feynman}
\end{figure}

Axions may be produced in pulsar cores through the process of nucleon-nucleon Bremsstrahlung as depicted in the Feynman diagram of Fig. \ref{fig:feynman}. The Bremsstrahlung process assumes a one pion exchange \textcolor{black}{(OPE)} approximation [\onlinecite{RN290}] and the nucleons involved are considered to be neutrons. Incoming nucleons \textit{N\textsubscript{1}, N\textsubscript{2}} and outgoing nucleons \textit{N\textsubscript{3}, N\textsubscript{4}} undergo one pion exchange to produce axions of energy $\omega$ via the Bremsstrahlung process. The axions then undergo radiative decay to gamma-ray photons.

The axion has a mass \textit{m\textsubscript{a}} which is related to the Peccei-Quinn scale \textit{f\textsubscript{a}} through a scaling relation (Eqn.~\ref{Scaling}). 

\begin{equation}{\label{Scaling}}
    m_{a} \approx 6\mu \, \mathrm{eV} \Big(\frac{f_{a}}{10^{12} \, \mathrm{GeV} }\Big)^{-1} 
\end{equation}

The spin structure function \textit{S\textsubscript{$\sigma$}($\omega$)} (Eqn.~\ref{SpinStructure}) is a phase space integral corresponding to the Bremsstrahlung process depicted in Fig.\ref{fig:feynman}.  The phase space integral accounts for nucleon spin and the balanced energy (\textit{E\textsubscript{1,2,3,4}}) and momenta (\textbf{p\textsubscript{1,2,3,4}}) transfer between nucleons \textit{N\textsubscript{1,2,3,4}} with conservation of momenta and energy provided by Dirac $\delta$ functions. The momenta \textbf{\textit{p\textsubscript{i}}} have integration limits in the range 0 \textless \textbf{\textit{p\textsubscript{i}}} \textless 2\textit{p\textsubscript{Fn}} where \textit{p\textsubscript{Fn}} is the neutron Fermi momentum. \textit{p\textsubscript{Fn}} is 300-400 MeV in supernovae cores [\onlinecite{RN219}] and typically \textgreater 100 MeV in neutron stars [\onlinecite{RN260}]. $\mathcal{F}$ in Eqn.~\ref{SpinStructure} is the product of thermodynamic functions as defined in Eqn.~\ref{ProductThermoDynamicFunctions}. $\mathcal{H}_{ij}$ is the hadronic tensor incorporating nucleon spin \textcolor{black}{with value  10/$\omega$\textsuperscript{2}. The rate of axion production can be determined independently of the OPE approximation using the soft-neutrino radiation rate which is proportional to the nucleon nucleon on-shell scattering amplitude. This soft-neutrino approximation (SNA) method gives an axion emission rate which is a factor of four smaller than that given by the OPE approximation [\onlinecite{RN260}]. It can be shown that a value of $\mathcal{H}_{ij}$ = 10/$\omega$\textsuperscript{2} largely includes the reduction in axion emission rate expected for the SNA by considering expressions for the scattering kernel of neutrinos produced by Bremsstrahlung in supernovae cores as presented in [\onlinecite{RN281}] where the SNA has not been applied. We can take the spin structure function  \textit{S\textsubscript{$\sigma$}($\omega$)} (Eqn.~\ref{SpinStructure}) to be analogous to the neutrino scattering kernel \textit{S\textsubscript{$\sigma$}($\omega$)} of [\onlinecite{RN281}] and thus equate $\mathcal{H}_{ij}$ to the spatial trace, $\overline{M}$, in the neutrino scattering kernel expression of [\onlinecite{RN281}]. By combining the expressions presented in [\onlinecite{RN281}] for a generic scattering kernel, the spin fluctuation rate and an effective degeneracy parameter, we obtain a $\mathcal{H}_{ij}$ value of 30/$\omega$\textsuperscript{2}. Thus, a value of 10/$\omega$\textsuperscript{2} for $\mathcal{H}_{ij}$ results in a factor of 3 reduction in axion emissivity which is comparable with the factor of 4 reduction expected from the SNA. } The thermodynamic function (Eqn.~\ref{ThermoDynamicFunction}) is the Fermi Dirac distribution \textcolor{black}{ in natural units (\textit{k\textsubscript{B}}=1)} for the nucleons applicable to degenerate matter [\onlinecite{RN213}] incorporating energy \textit{E}, temperature \textit{T} and neutron star degeneracy $\mu$. We take the value of $\mu$/T = 10 as used in the analysis of [\onlinecite{RN131}].

\begin{equation}{\label{SpinStructure}}
    \begin{aligned}
    S\textsubscript{$\sigma$}(\omega)= \frac{1}{4}\int\Big[\prod_{i=1..4}
    \frac{d^3p\textsubscript{i}}{(2\pi)^3}\Big] \\
    \times\, (2\pi)^4\delta^3(\mathbf{p_{1}}+\mathbf{p_{2}}-\mathbf{p_{3}}-\mathbf{p_{4}}) \\
    \times\, \delta(E_{1}+E_{2}-E_{3}-E_{4}-\omega) \mathcal{F} \mathcal{H}\textsubscript{ij}
    \end{aligned} 
\end{equation}

\begin{equation}{\label{ProductThermoDynamicFunctions}}
    \mathcal{F}= f(E_{1})f(E_{2})(1-f(E_{3})(1-f(E_{4}))
\end{equation}

\begin{equation}{\label{ThermoDynamicFunction}}
    f(E)=1/(1+exp((E-\mu )/T))
\end{equation}

The axion emissivity or energy loss rate per volume in natural units (i.e. \textit{$\hbar$}=\textit{c}=1), $\epsilon_a$ is defined by Eqn.~\ref{AxionEmissivity} as given in [\onlinecite{RN260}] where \textit{M\textsubscript{N}} is the nucleon mass of 938 MeV and g\textsubscript{ann} is the axion-nucleon coupling with \textit{g\textsubscript{ann}}= \textit{C\textsubscript{N}}\textit{M\textsubscript{N}}/\textit{f\textsubscript{a}}. \textit{C\textsubscript{N}} encapsulates the vacuum expectation values for the Higgs \textit{u} and \textit{d} doublets with the doublets giving mass to the up and down quarks of the nucleons. The value of \textit{C\textsubscript{N}} depends on the coupling model considered with 0 \textless\textit{C\textsubscript{N}} \textless 2.93 [\onlinecite{MAYLE1988188}]; we take \textit{C\textsubscript{N}}=0.1 as [\onlinecite{RN131}].

\begin{equation}{\label{AxionEmissivity}}
    \epsilon_{a}= \frac{g^2\textsubscript{ann}}{48\pi^2M_{N}^2}\int \omega^4S_{\sigma}(\omega) \, d\omega
\end{equation}

The expected photon flux arising from axion decay for a photon of energy $E$ is given by Eqn.~\ref{PhotonFlux} from [\onlinecite{RN131}] where \textit{d} is the distance to the pulsar in parsecs and $\Delta$\textit{t} is the timescale for the emission of axions from a neutron star with a core temperature of 20 MeV (Eqn.~\ref{MeanFreeTime}). We take the value of \textit{S\textsubscript{$\sigma$}($\omega$)} to be 2.4 $\times$ 10\textsuperscript{7} MeV\textsuperscript{2} and 6.25 $\times$ 10\textsuperscript{4} MeV\textsuperscript{2} for axion energies of 100 MeV and 200 MeV respectively from the values of \textit{$\omega$\textsuperscript{4}S\textsubscript{$\sigma$}($\omega$)} in the axion emissivity versus energy plot of [\onlinecite{RN131}] for a pulsar of core temperature 20 MeV and $\mu$/T = 10. We choose \textit{S\textsubscript{$\sigma$}($\omega$)} at $\omega$=100 MeV and $\omega$=200 MeV in our calculations  because these represent reasonable extremes on the emissivity plot, with emissivity peaking and being less sensitive to energy near $\omega$=100 MeV and an emissivity cut-off at $\omega$=230 MeV. 

\begin{equation}{\label{PhotonFlux}}
    \begin{aligned}
    E \frac{d\Phi}{dE}= 1.8 \times 10^{-2} \Big (\frac{m_a}{\mathrm{eV}} \Big )^5 \Big( \frac{\Delta \, t}{23.2 \, \mathrm{s}}  \Big ) \Big( \frac{100 \, \mathrm{pc}}{d}  \Big )^2 \\
    \times\, \Big ( \frac {2E}{100 \, \mathrm{MeV} }   \Big)^4 \Big( \frac{S_\sigma (2E)}{10^7 \, \mathrm{MeV^2}}  \Big) \, \mathrm{cm^{-2} \, s^{-1}}
    \end{aligned}
\end{equation}

\begin{equation}{\label{MeanFreeTime}}
    \Delta t=23.2\, s \,  \Big (\frac{\mathrm{eV}}{m_a} \Big )^2 
\end{equation}

By combining Eqn.~\ref{PhotonFlux} and Eqn.~\ref{MeanFreeTime} the UL axion mass can be expressed in terms of the UL gamma-ray photon flux $\Phi$ of a pulsar (Eqn.~\ref{AxionMassFromPhotonFlux}).

\begin{equation}{\label{AxionMassFromPhotonFlux}}
    \begin{aligned}
    \mathrm{UL} \,  m_a= \Big[ \mathrm{UL \,} \Phi \, \mathrm{cm^{-2} \, s^{-1}}  \, \times\, 55.5 \, \times\, \Big ( \frac {d}{100 \, \mathrm{pc}}  \Big )^2  \\ \times \, \Big ( \frac{100 \, \mathrm{MeV}}{2E}\Big)^4 \Big ( \frac{10^7 \, \mathrm{MeV}^2}{S_\sigma (2E)} \Big ) \Big]^{\frac{1}{3}}
    \end{aligned}
\end{equation}

Alternatively, instead of using photon flux methods as described above, axion mass can be constrained using an expression for the energy lost from the pulsar as a result of axion production. The energy loss rate \textit{$\epsilon$\textsubscript{a}\textsuperscript{D}} for a given mass of neutron star material arising from the production of axions in the pulsar core (Eqn.~\ref{Raffelt}) is as presented in [\onlinecite{RAFFELT1996}] based on [\onlinecite{RN290}] and [\onlinecite{RN213}] with $\alpha$\textsubscript{a} as Eqn.~\ref{alphaA}. \textit{T}\textsubscript{MeV} is the neutron star core temperature in MeV and $\rho$\textsubscript{15} is the neutron star mass density in units of 10\textsuperscript{15} g cm\textsuperscript{-3} \textcolor{black}{We include a further factor of 0.25 in Eqn.~\ref{Raffelt} to allow for the SNA reduction in axion emission rate}. 

\begin{equation}{\label{Raffelt}}
    \begin{aligned}
     {\epsilon_a}^D = 0.25  \times \alpha_a 1.74 \times 10^{31} \mathrm{erg \, g^{-1} \, s^{-1}} \, \rho_{15}^{-2/3} T_{MeV}^6
    \end{aligned}
\end{equation}

\begin{equation}{\label{alphaA}}
    \begin{aligned}
     {\alpha_a} \equiv \Big(\frac{C_N M_N}{f_a}\Big)^2/4\pi
    \end{aligned}
\end{equation}

The measured UL gamma-ray luminosity, \textit{L\textsubscript{$\gamma$}} can be equated to the expected gamma-ray luminosity arising from the axion energy loss rate for the total mass of the neutron star as \textit{ L\textsubscript{$\gamma$}}=\textit{$\epsilon$\textsubscript{a}\textsuperscript{D}NS\textsubscript{mass} P\textsubscript{$a\rightarrow \gamma$}}, where \textit{NS\textsubscript{mass}} is the neutron star mass expressed in grams and \textit{P\textsubscript{$a\rightarrow \gamma$}} is the axion to photon conversion probability (0-1.0) in the pulsar \textit{B} field. In the case of axion radiative decay where an axion decays to two gamma-ray photons, \textcolor{black} {without conversion in the pulsar B field being required, we take}    \textit{P\textsubscript{$a\rightarrow \gamma$}} \textcolor{black}{to be} 1.1 $\times$ 10\textsuperscript{-24} s\textsuperscript{-1}(\textit{m\textsubscript{a}}/1 eV) \textsuperscript{5}\textcolor{black}{ [\onlinecite{RN341}].} From the above expression for \textit{L\textsubscript{$\gamma$}} and by combining Eqns.~\ref{Scaling}, ~\ref{Raffelt} and~\ref{alphaA} we obtain an expression for UL \textit{m\textsubscript{a}} (Eqn.~\ref{UL_Axion_Mass}). We assume a canonical pulsar mass of 1.4 M\textsubscript{$\odot$} or 2.786 $\times$ 10\textsuperscript{33} g and a density of 0.056 $\times$ 10\textsuperscript{15} g cm\textsuperscript{-3}.

        \begin{equation*}{\label{UL_Axion_Mass}}
        \begin{aligned}
        \mathrm{UL} \,  m_a= \frac{6.0 \times 10^{15}}{C_N M_N}   \times
         \end{aligned}
         \end{equation*}
         
        \begin{equation}{\label{UL_Axion_Mass}}
        \begin{aligned}
          \Big( \frac{4\pi L_\gamma \, \mathrm {erg \, s^{-1}}}{0.435 \times 10^{31} \mathrm{erg \, g^{-1} \, s^{-1}} \, \rho_{15}^{-2/3} T_{MeV}^6 NS  _\mathrm{mass} P\textsubscript{$a\rightarrow \gamma$}  }  \Big)^{\frac{1}{2}}
         \end{aligned}
         \end{equation}

\section{Pulsar Selection}
\label{sec:GCSelection}
We make the simple assumption that axions are emitted in a continuous isotropic fashion by the pulsar and are unaffected by pulsar rotation. In making our pulsar selection we want to maximise the probability of detecting isotropic gamma-ray emission arising solely from the decay of axions to gamma-rays. Thus we wish to exclude the pulsed gamma-ray emission arising from pulsar magnetospheric emission which would be unrelated to axion production and a background to the axion signal that we wish to measure. Therefore, our selection of 17 pulsars (Table~\ref{tab:psr_list}) from version 1.57 of the Australia Telescope National Facility(ATNF) catalogue[\onlinecite{ATNF1}] \footnote{http://www.atnf.csiro.au/research/pulsar/psrcat/ } is based on the following criteria to minimise gamma-ray background and to select well-measured pulsars which are most likely to emit detectable gamma-rays solely through axion decay: 

\begin{itemize}
\item We include pulsars which are located off the Galactic plane (\textbar \textit{b}\textbar\textgreater 15\textdegree{}) thus reducing the uncertainty arising from the Galactic gamma-ray background model of the Galactic disc
\item We include pulsars away from the Galactic centre with \textit{l}\textgreater 30\textdegree{} and \textit{l}\textless 330\textdegree{}
\item We include nearby pulsars with a heliocentric distance of 0.5 kpc or less and possessing an $\dot{E}$ \textgreater 0 in the ATNF catalogue
\item We include only pulsars which are not known to have binary companions in the ATNF catalogue and have not been identified as prior sources of gamma-ray emission in either the Public List of LAT-Detected Gamma-Ray Pulsars\footnote{https://confluence.slac.stanford.edu/display/GLAMCOG/ \\ Public+List+of+LAT-Detected+Gamma-Ray+Pulsars, list last updated 19\textsuperscript{th} Oct  2018, accessed on 14\textsuperscript{th} Feb 2019} (which lists all publicly-announced gamma-ray pulsar detections, whose significance exceeds 4$\sigma$) or in the Second \textit{Fermi} Large Area Telescope Catalog of Gamma-Ray Pulsars [\onlinecite{RN244}].

\end{itemize}

\begin{table*}
	\centering

	\begin{tabular}{c c c c c c c c c c c c}
        \hline
        Name&&\textit{l} &\textit{b}&RA&Dec&Period (s)&Distance&$B$ Surface&$B$ Light&$\dot{E}$&Spin Down\\
        and Ref.&&(degree)&(degree)&(degree)&(degree)&and Ref.&(kpc)&(10\textsuperscript{10} Gauss)&Cylinder (Gauss)&(10\textsuperscript{30} erg s\textsuperscript{-1})&Age (10\textsuperscript{5} Yr) \\
        \hline

J0736-6304 & [\onlinecite{bb10}]	&	274.88	&	-19.15	&	114.08	&	-63.07	&	4.863 [\onlinecite{jcs+17}] 	&	0.10	&	2750.00	&	2.24	&	52.1&5.07	\\
J0711-6830 &[\onlinecite{bjb+97}] &   279.53  &   -23.28  &   107.98  &   -68.51  &   0.005 [\onlinecite{rhc+16}]  &   0.11    &   0.03 & 16400 & 3550& 58400\\
J0536-7543 &[\onlinecite{mlt+78}]&	287.16	&	-30.82	&	84.13	&	-75.73	&	1.246 [\onlinecite{smd93}]	&	0.14	&	84.90	&	4.12	&	11.5&349	\\
J0459-0210 &[\onlinecite{mld+96}]	&	201.44	&	-25.68	&	74.97	&	-2.17	&	1.133 [\onlinecite{hlk+04}]	&	0.16	&	127.00	&	8.21	&	37.9&128	\\
J0837+0610 &[\onlinecite{phbc68}]	&	219.72	&	26.27	&	129.27	&	6.17	&	1.274 [\onlinecite{hlk+04}]	&	0.19	&	298.00	&	13.50	&	130.0&29.7	\\
J0108-1431 &[\onlinecite{tnj+94}]&	140.93	&	-76.82	&	17.03	&	-14.53	&	0.808 [\onlinecite{hlk+04}]	&	0.21	&	25.20	&	4.49	&	5.8&1660	\\
J0953+0755 &[\onlinecite{phbc68}]	&	228.91	&	43.70	&	148.29	&	7.93	&	0.253 [\onlinecite{hlk+04}]	&	0.26	&	24.40	&	141.00	&	560.0&175	\\
J1116-4122 &[\onlinecite{mlt+78}]	&	284.45	&	18.07	&	169.18	&	-41.38	&	0.943 [\onlinecite{antt94}]	&	0.28	&	277.00	&	31.00	&	374.0&18.8	\\
J0630-2834 &[\onlinecite{lvw69a}]	&	236.95	&	-16.76	&	97.71	&	-28.58	&	1.244 [\onlinecite{hlk+04}]	&	0.32	&	301.00	&	14.70	&	146.0&27.7	\\
J0826+2637 &[\onlinecite{cls68}]	&	196.96	&	31.74	&	126.71	&	26.62	&	0.531 [\onlinecite{hlk+04}]	&	0.32	&	96.40	&	60.50	&	452.0&49.2	\\
J1136+1551 &[\onlinecite{phbc68}]	&	241.90	&	69.20	&	174.01	&	15.85	&	1.188 [\onlinecite{hlk+04}]	&	0.35	&	213.00	&	11.90	&	87.9&50.4	\\
J0656-5449 &[\onlinecite{jbo+09}]	&	264.80	&	-21.14	&	104.20	&	-54.82	&	0.183 [\onlinecite{jbo+09}]	&	0.37	&	7.74	&	118.00	&	205.0&909	\\
J0709-5923 &[\onlinecite{jbo+09}]	&	270.03	&	-20.90	&	107.39	&	-59.40	&	0.485 [\onlinecite{jbo+09}]	&	0.37	&	25.00	&	20.50	&	43.5&610	\\
J0636-4549 &[\onlinecite{bjd+06}]	&	254.55	&	-21.55	&	99.14	&	-45.83	&	1.985 [\onlinecite{bjd+06}]	&	0.38	&	254.00	&	3.05	&	16.0&99.1	\\
J0452-1759 &[\onlinecite{vlw69}]	&	217.08	&	-34.09	&	73.14	&	-17.99	&	0.549 [\onlinecite{hlk+04}]	&	0.40	&	180.00	&	102.00	&	1370.0&15.1	\\
J0814+7429 &[\onlinecite{cp68}]	&	140.00	&	31.62	&	123.75	&	74.48	&	1.292 [\onlinecite{hlk+04}]	&	0.43	&	47.20	&	2.05	&	3.1&1220	\\
J2307+2225 &[\onlinecite{cnt96}]	&	93.57	&	-34.46	&	346.92	&	22.43	&	0.536 [\onlinecite{cn95}]	&	0.49	&	6.91	&	4.21	&	2.2&9760	\\

		\hline
	\end{tabular}
    	\caption{Our selection of 17 pulsars from the ATNF catalogue showing their Galactic longitude/latitude, RA and Dec co-ordinates, period, pulsar distance, magnetic field $B$ at surface and light cylinder in Gauss, $\dot{E}$ and spin down age. Discovery and period are from the references listed. }
        \label{tab:psr_list}
\end{table*}

\section{Analysis}
\label{sec:Analysis}
\subsection{Photon Event Data Selection}
\label{sec:PhotonEventDataSelection} 
The data in this analysis were collected by \textit{Fermi}-LAT between 4th Aug 2008 to 18th October 2017 (Mission Elapsed Time (MET)  2395574147[s] to 530067438[s]). We consider all \textsc{pass} 8 events which are \textit{source} class  photons (evclass=128), with Front converting events (evtype=1), spanning the energy range 60 to 500 MeV. We use Front\textcolor{black}{\footnote{\textcolor{black}{We have repeated the same analysis using the PSF3 event class which is the best quartile direction reconstruction. This does not
change the determined \textit{m\textsubscript{a}} significantly considering all 17 PSRs. We therefore retain the FRONT analysis to allow direct comparison with [\onlinecite{RN131}].} }} converting events because of the improved point spread function (PSF) of this event class with 95 per cent containment of 60 MeV photons at a containment angle of 13\textdegree{} as opposed to 20\textdegree{} for both Front and Back converting events. We select a conservative energy range of 60-500 MeV, as axion decay has previously been expected to produce gamma-rays in the range 60-200 MeV, with a cut-off by 200 MeV [\onlinecite{RN131}]. Throughout our analysis, the \textit{Fermipy} software package\footnote{\textit{Fermipy} change log version 0.12.0 }[\onlinecite{2017arXiv170709551W}] with version \textsc{v10r0p5} of the \textit{Fermi Science Tools} is used, in conjunction with the \textsc{p8r2\_source\_v6} instrument response functions. We apply the standard \textsc{pass} 8 cuts to the data, including a zenith angle 90\textdegree{} cut to exclude photons from the Earth limb and good-time-interval cuts of DATA\_QUAL \textgreater 0 and LAT\_CONFIG = 1. The energy binning used is 4 bins per decade in energy and spatial binning is 0.1\textdegree{} per image pixel.

\subsection{Determining if Pulsars are Gamma-ray Emitters }
\label{sec:InititialDetectionGC} 

We first determine if any of the pulsars in our selection are significant unpulsed gamma-ray emitters. For each pulsar we consider a 20\textdegree{} Radius of Interest (ROI) centred on the pulsar co-ordinates. We use an ROI of 20\textdegree{} as our analysis is made down to a low energy of 60 MeV and we wish to be certain to allow for the contribution of low energy sources given the PSF of 13\textdegree{} above.

We include known sources using a point source population derived from the \textit{Fermi}-LAT's third point source catalog (3FGL), diffuse gamma-ray emission and extended gamma-ray sources. The diffuse gamma-ray emission consists of two components: the Galactic diffuse flux and the isotropic diffuse flux. The Galactic component is modelled with \textit{Fermi}-LAT's gll\_iem\_v06.fit spatial map with the normalisation free to vary. The isotropic diffuse emission is defined by \textit{Fermi's}  iso\_\textsc{P8R2}\_\textsc{SOURCE}\_\textsc{V6}.txt tabulated spectral data. The normalisation of the isotropic emission is also left free to vary. In addition, all known sources take their spectral shape as defined in the 3FGL catalogue.

An energy dispersion correction is applied to the pulsar test source but disabled for all 3FGL sources in line with Fermi Science Support Centre recommendations for low energy analysis.

We perform an initial \textsc{binned} likelihood analysis using the \textit{optimize} method with the normalisation of all point sources within 20 \textdegree{} of the pulsar being left free.  

From this initial likelihood fit, all point sources (with the exception of the target pulsar) with a TS $<4$, or with a predicted number of photons, $Npred$ $<4$ are removed from the model. Thereafter, we free the spectral shape of all TS $>25$ sources in this refined model and undertake a further secondary likelihood fit using \textit{optimize} and \textit{fit} methods.

The best-fit model from this secondary likelihood fit is then used with the \textit{Fermi Science Tool} \textsc{gttsmap}, to search for new point sources that were not already present in the 3FGL. In particular, we run \textit{Fermipy's `find\_sources'} method to detect all sources above 3$\sigma$ significance. \textit{Find\_sources} is a peak detection algorithm which analyses the test statistic (TS) map to find new sources over and above those defined in the 3FGL model by placing a test point source, defined as a power law  with spectral index 2.0, at each pixel on the TS map and recomputing likelihood. Lastly, we again run the \textit{fit} method to perform a final likelihood fit, which fits all parameters that are currently free in the model and updates the TS and predicted count ($Npred$) values of all sources. 

\subsection{Pulsar Upper Limit Gamma-ray Emission }
\label{sec:PSRULMETHOD} 

In order to determine PSR gamma-ray flux upper limits we repeat the analysis of Section \ref{sec:InititialDetectionGC} with a source model which includes a pulsar test source for each of the 17 pulsars. The differential flux, \textit{dN/dE}, (photon flux per energy bin) of the test source for each pulsar is described as a power law \footnote{As described in the Fermi Science Support Centre link  https\://fermi.gsfc.nasa.gov/ssc/data/analysis/scitools/source\_models.html} as defined in Eqn.~\ref{PLeqn} where \textit{prefactor} = $N_{0}$, \textit{index}=$\gamma$ and \textit{scale}=$E_{0}$. The test source has \textit{index} of 2.0, a \textit{scale} of 1 GeV and a \textit{prefactor} = 1 $\times$ 10 \textsuperscript{-11}. \textcolor{black}{We leave the \textit{prefactor} (normalisation) and \textit{index} of the test source free to vary.} 

\begin{equation}{\label{PLeqn}}
\frac{dN}{dE}=N\textsubscript{0}\Big(\frac{E}{E\textsubscript{0}}\Big)^\gamma\
\end{equation}

We then obtain UL photon and energy fluxes integrated over the energy analysis range (at 2 $\sigma$ significance, 95 percent confidence level) from the flux\_ul95 and eflux\_ul95 attributes respectively of the  \textit{fermipy} sources entry for each pulsar test source. The UL photon and energy fluxes are defined as the values where the likelihood function, 2$\Delta$Log(L), which compares the likelihood of a model with the source and without, has decreased by 2.71 from its maximum value across the range of flux values arising from the analysis. \textcolor{black}{In addition, we use a composite likelihood stacking technique to improve the UL photon flux determination by considering all test sources in the analysis together. We extract a likelihood profile of $\Delta$Log(L) vs photon flux for each test source using the  \textit{fermipy profile\_norm} method. Next we determine the functional form of this likelihood profile for each test source using \textit{numpy polyfit} and \textit{poly1d} and interpolate the likelihood profile with \textit{numpy polyval} between the overall minimum and maximum photon flux value obtained by considering the UL photon flux of all test sources. We then sum the $\Delta$Log(L) values of each interpolated likelihood profile to obtain a single stacked  $\Delta$Log(L) vs photon flux profile for the test sources as a whole. Finally, we determine the maximum photon flux where the stacked $\Delta$Log(L) has decreased by 1.35 from its peak value to give the one-sided upper limit photon flux.}

\section{Results}
\label{sec:Results}

\begin{table*}
	\centering

	\begin{tabular}{c c c c c c c}
        \hline
        Pulsar   &      TS      & UL Photon Flux                                                         & UL Energy Flux  & UL $\gamma$ Luminosity & UL \textit{m\textsubscript{a}}  $\omega$=100 MeV & UL \textit{m\textsubscript{a}} $\omega$=200 MeV   \\
                &              & (10\textsuperscript{-8} cm\textsuperscript{-2} s\textsuperscript{-1}) & (10\textsuperscript{-12} erg cm\textsuperscript{-2} s\textsuperscript{-1}) & (10\textsuperscript{31} erg s\textsuperscript{-1}) &   (10\textsuperscript{-2} eV)    &  (10\textsuperscript{-2} eV)       \\
        \hline

J0711-6830	&	3	&	0.04	&	1.51	&	0.22	&	0.21	&	0.70	\\
J0536-7543	&	0	&	0.22	&	0.53	&	0.12	&	0.43	&	1.45	\\
J0837+0610	&	0	&	0.27	&	0.63	&	0.27	&	0.57	&	1.90	\\
J0108-1431	&	0	&	0.18	&	0.41	&	0.21	&	0.52	&	1.75	\\
J0953+0755	&	2	&	0.47	&	1.32	&	1.07	&	0.84	&	2.81	\\
J1116-4122	&	1	&	0.90	&	1.73	&	1.62	&	1.09	&	3.66	\\
J0826+2637	&	2	&	0.39	&	1.18	&	1.44	&	0.91	&	3.04	\\
J1136+1551	&	0	&	0.50	&	1.16	&	1.70	&	1.04	&	3.49	\\
J0656-5449	&	0	&	0.32	&	0.75	&	1.23	&	0.94	&	3.14	\\
J0636-4549	&	3	&	1.31	&	2.08	&	3.60	&	1.52	&	5.08	\\
J0452-1759	&	0	&	0.31	&	0.71	&	1.36	&	0.97	&	3.24	\\
J0814+7429	&	0	&	0.23	&	0.54	&	1.19	&	0.93	&	3.10	\\

		\hline
	\end{tabular}
    	\caption{Test statistic, UL photon flux, UL energy flux, UL gamma luminosity and UL \textit{m\textsubscript{a}} for axion energies of 100 and 200 MeV for the 12 undetected pulsars. }
        \label{tab:psr_results}
\end{table*}

\begin{table*}
	\centering

	\begin{tabular}{c c c c c c c}
        \hline
        Pulsar   &      TS      & UL Photon Flux                                                         & UL Energy Flux  & UL $\gamma$ Luminosity & UL \textit{m\textsubscript{a}}  $\omega$=100 MeV & UL \textit{m\textsubscript{a}} $\omega$=200 MeV   \\
                &              & (10\textsuperscript{-8} cm\textsuperscript{-2} s\textsuperscript{-1}) & (10\textsuperscript{-12} erg cm\textsuperscript{-2} s\textsuperscript{-1}) & (10\textsuperscript{31} erg s\textsuperscript{-1}) &   (10\textsuperscript{-2} eV)    &  (10\textsuperscript{-2} eV)       \\
        \hline

J0736-6304	&	33	&	2.68	&	4.87	&	0.58	&	0.79	&	2.65	\\
J0459-0210	&	10	&	1.72	&	3.64	&	1.11	&	0.93	&	3.13	\\
J0630-2834	&	19	&	1.89	&	3.59	&	4.40	&	1.53	&	5.12	\\
J0709-5923	&	12	&	1.03	&	2.55	&	4.17	&	1.38	&	4.62	\\
J2307+2225	&	14	&	1.12	&	2.87	&	8.25	&	1.71	&	5.72	\\

		\hline
	\end{tabular}
    	\caption{Test statistic, UL photon flux, UL energy flux, UL gamma luminosity and UL \textit{m\textsubscript{a}} for axion energies of 100 and 200 MeV for the 5 pulsars which are associated with areas of extended diffuse gamma-ray emission. }
        \label{tab:psr_results_detected}
\end{table*}

\subsection{Pulsar UL Gamma-ray Fluxes}

 We list the UL photon, energy fluxes and  gamma-ray luminosities (assuming the distances in Table~\ref{tab:psr_list}) for our sample of pulsars in Tables~\ref{tab:psr_results} and~\ref{tab:psr_results_detected}. \textcolor{black}{The UL photon flux at 95 percent confidence obtained by likelihood stacking of all 17 pulsars is 7.8 $\times$ 10\textsuperscript{-10} cm\textsuperscript{-2} s\textsuperscript{-1}. }

\subsection{Upper Limit \textit{m\textsubscript{a}} Determination}
\label{sec:result_ul_ma}

We list our determination of UL \textit{m\textsubscript{a}} in Tables~\ref{tab:psr_results} and~\ref{tab:psr_results_detected} for each pulsar derived from the UL photon flux and Eqn.~\ref{AxionMassFromPhotonFlux} for axions of energy 100 MeV and 200 MeV. The average UL \textit{m\textsubscript{a}} considering all 17 pulsars is 9.6 $\times$ 10\textsuperscript{-3} eV and 3.21 $\times$ 10\textsuperscript{-2} eV for axions of energy 100 MeV and 200 MeV respectively. We obtain an average UL \textit{m\textsubscript{a}} for the 4 pulsars analysed in [\onlinecite{RN131}], J0108-1431, J0953+0755, J0630-2834 and J1136+1551 of 9.8 $\times$ 10\textsuperscript{-3} eV and 3.29 $\times$ 10\textsuperscript{-2} eV for axions of energy 100 MeV and 200 MeV respectively. 

Our determination of UL \textit{m\textsubscript{a}} = 9.6 $\times$ 10\textsuperscript{-3} eV is a factor of 8 improvement on the result of [\onlinecite{RN131}] who determined an UL \textit{m\textsubscript{a}} of 7.9 $\times$ 10\textsuperscript{-2} eV.

\textcolor{black}{Finally, we note that the UL \textit{m\textsubscript{a}} obtained by likelihood stacking is improved two-fold compared to the averaged result above, with UL \textit{m\textsubscript{a}} of 4.8 $\times$  10\textsuperscript{-3} eV and 1.61 $\times$ 10\textsuperscript{-2} eV for axions of energy 100 MeV and 200 MeV respectively.}

\subsection{Pulsars Near Extended Emission}

\begin{figure*}
	\includegraphics[width=15cm,keepaspectratio]{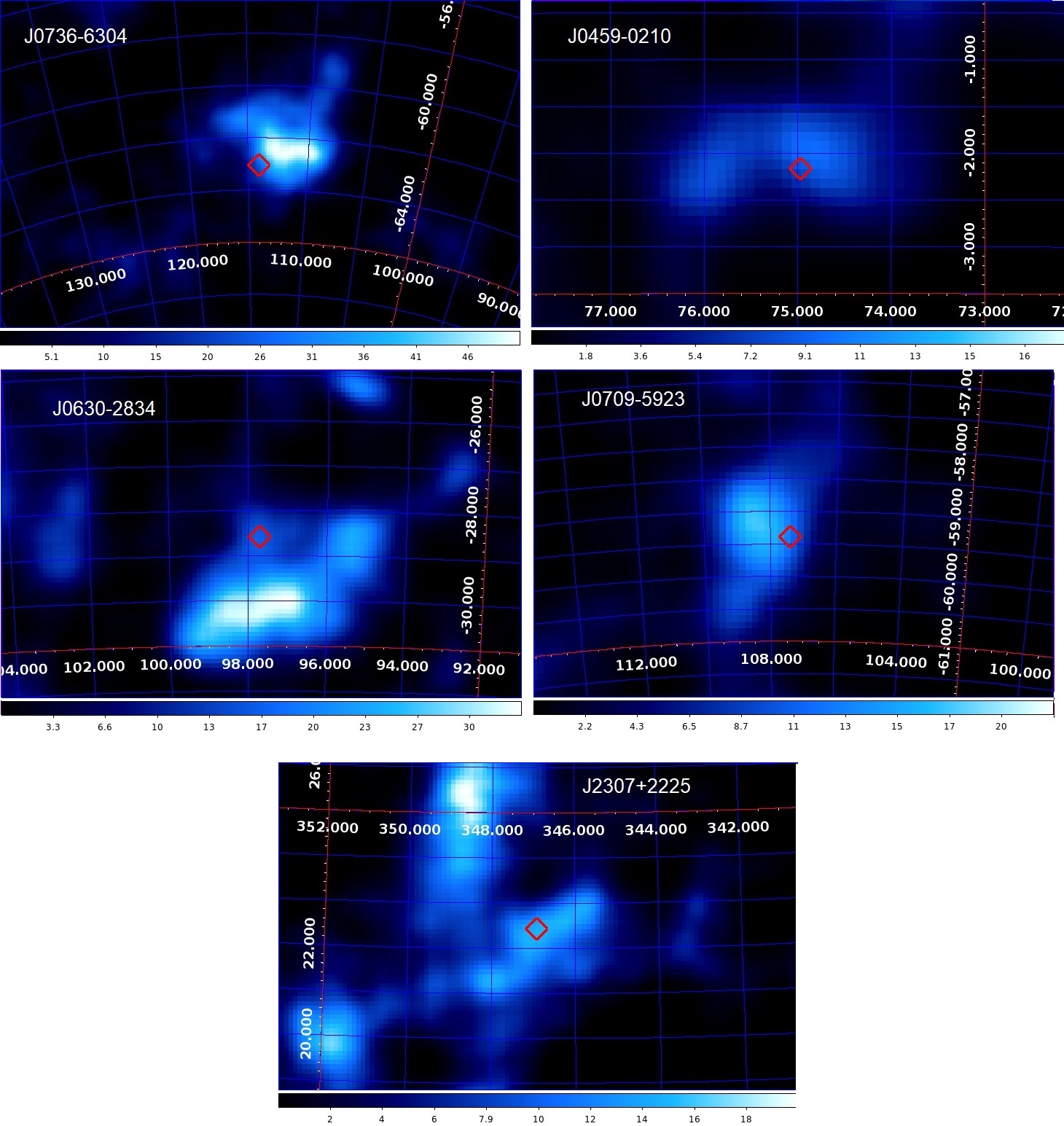}
    \caption{TS maps for our gamma-ray analysis of the 5 pulsar test sources detected at \textgreater 3 $\sigma$ significance (Table~\ref{tab:psr_results_detected}) showing that these sources are inconsistent with a point source detection characteristic of pulsars and part of extended diffuse features. The horizontal contour scale is the TS value, the red diamond is the pulsar position, horizontal axis is RA in decimal degrees, vertical axis is Dec in decimal degrees. }
    \label{fig:COMBINED_TS_MAP}
\end{figure*}

  We note that the UL test sources for 5 pulsars are detected with a significance which exceeds 3 $\sigma$ , namely J0736-6304 5.7 $\sigma$ (TS 33), J0630-2834 4.4 $\sigma$ (TS 19), J2307+2225 3.7 $\sigma$ (TS 14),  J0709-5923 3.5 $\sigma$ (TS 12) and J0459-0210 3.2 $\sigma$ (TS 10). However, the initial analysis which searches for point sources (whilst not introducing a pulsar test source), detects no point sources at the pulsar co-ordinates and thus we discount these apparent detections as true detections of the pulsars concerned. The lack of significant point source pulsar detections can also be seen on TS maps for the analysis (Fig.~\ref{fig:COMBINED_TS_MAP}) where the pulsars are spatially co-incident with regions of extended gamma-ray emission uncharacteristic of the point source emission expected from a pulsar. 
  
  We also check for source extension of the pulsars by running the GTAnalysis \textit{extension} method. \textit{extension} replaces the pulsar point source spatial model with an azimuthally symmetric 2D Gaussian model. It then profiles likelihood with respect to spatial extension in a 1 dimensional scan to determine the likelihood of extension.  Only the J0736-6304 test source has some evidence of extension with an extension TS value of 14 (3.7 $\sigma$). The remaining 4 pulsars with significance \textless 4.4 $\sigma$ are consistent with background and as expected have no significant extension.

\textcolor{black}{We make the assumption that axion emission is isotropic and so the extended emission of J0736-6304 which is asymmetric and exhibits its highest significance offset from the pulsar would seem to be inconsistent with an axion source. Instead, this emission is more likely to be consistent with variations in the Galactic diffuse gamma-ray background. }

  These 5 pulsars generally exhibit higher UL fluxes (Table ~\ref{tab:psr_results_detected}) than the other 12 (Table ~\ref{tab:psr_results}) and so omitting these 5 pulsars from the determination of UL \textit{m\textsubscript{a}} yields an improved average UL \textit{m\textsubscript{a}} for the 12 remaining pulsars of 8.9 $\times$ 10\textsuperscript{-3} eV and 2.97 $\times$ 10\textsuperscript{-2} eV for axions of energy 100 MeV and 200 MeV respectively.

\section{Discussion}
\label{sec:Discussion}

\subsection{Upper Limit Determination}
\label{sec:discussion_ul_ma}

The authors of [\onlinecite{RN131}] analysed 4 pulsars J0108-431, J0953+0755, J0630-2834 and J1136+1551 with an \textsc{unbinned} likelihood analysis using the 2FGL catalogue, 5 years of \textit{Fermi}-LAT \textsc{pass} 7 event data in the energy range 60$-$200 MeV and employing front converting source photon events. They detected no gamma-ray emission and determined a 95 percent confidence UL photon flux for each of the 4 pulsars using the \textsc{minos} method of the \textit{Fermi} Science Tools. In contrast, we analyse 17 pulsars (including the 4 pulsars of [\onlinecite{RN131}]) with a \textsc{binned} likelihood analysis using the 3FGL catalogue and 9 years of \textit{Fermi}-LAT \textsc{pass} 8 event data in the energy range 60$-$500 MeV, again using front converting events. We determine the UL photon flux using the \textit{fermipy} flux\_ul95 entry for each pulsar. Using this analysis we obtain UL photon fluxes (Table~\ref{tab:berenji_fluxes}) comparable to [\onlinecite{RN131}] for the 4 pulsars they consider, which serves as a useful check of our gamma-ray analysis method, and do not detect any pulsars in our sample. 

Our method to determine  UL \textit{m\textsubscript{a}} differs from [\onlinecite{RN131}] in that we use UL photon fluxes directly as input to Eqn.~\ref{AxionMassFromPhotonFlux} whilst they fit a model of the spectral energy distribution (SED) of differential flux to a stacked likelihood analysis of the 4 pulsars using the \textit{COMPOSITE2} module of the \textit{Fermi} science tools and take the UL normalisation  of this model to be UL (\textit{m\textsubscript{a}} /eV)\textsuperscript{5} from which  they obtain UL \textit{m\textsubscript{a}} with all flux dependencies on astrophysical factors being accounted for in the SED model. 

We can use the UL photon fluxes obtained by [\onlinecite{RN131}] to consider the improvement in UL \textit{m\textsubscript{a}} determination which arises from our UL \textit{m\textsubscript{a}} calculation method alone. The average UL \textit{m\textsubscript{a}} for the 4 pulsars using the [\onlinecite{RN131}] photon fluxes (Table~\ref{tab:berenji_fluxes}) and our method (Eqn.~\ref{AxionMassFromPhotonFlux}) is 9.7 $\times$ 10\textsuperscript{-3} eV and 3.25 $\times$ 10\textsuperscript{-2} eV for axions of energy 100 MeV and 200 MeV, improving on the 7.9 $\times$ 10\textsuperscript{-2} eV determination of [\onlinecite{RN131}] by a factor of 2.4$-$8.1. Despite this improvement, we note that our determination of UL \textit{m\textsubscript{a}} is conservative because we assume that the integrated UL photon flux arises solely from a specific axion energy (100 MeV or 200 MeV) rather than the lower UL flux (and hence more constraining) UL \textit{m\textsubscript{a}} determination  which would be expected if we could determine UL photon flux for each energy bin in the analysis energy range of 60$-$500 MeV.     

We determine a very similar UL \textit{m\textsubscript{a}} in our sample of 17 pulsars of 9.6 $\times$ 10\textsuperscript{-3} eV and 3.21 $\times$ 10\textsuperscript{-2} eV for axions of energy 100 MeV and 200 MeV respectively. \textcolor{black}{These results are also comparable with  UL \textit{m\textsubscript{a}} values  obtained by modelling the cooling of Cassiopeia A observed by \textit{Chandra}. By assuming that the cooling results from both neutrino and axion emission and that a state of superfluidity exists in the star, an UL \textit{m\textsubscript{a}} of (1.7 $-$ 4.8)  $\times$ 10\textsuperscript{-2} eV  is obtained for \textit{C\textsubscript{N}} = ( 0.14 $-$ -0.05 )[\onlinecite{RN342}].}

As a final check to test whether the SED differential flux model used by [\onlinecite{RN131}] can be fitted individually to any of our 17 pulsars, we add a test source with the SED differential flux model from [\onlinecite{RN131}] implemented using the \textit{FileFunction} spectral model (Eqn.~\ref{FileFiteqn}) with flux values as Table~\ref{tab:FileFunction}  and re-analyse as Section \ref{sec:Analysis} above. All 17 pulsars remain undetected with the differential flux model test source exhibiting a  consistent normalisation of 10 \textsuperscript{-5} for all pulsars which is equivalent to  \textit{m\textsubscript{a}} \textless 0.1 eV.

\begin{equation}{\label{FileFiteqn}}
\frac{dN}{dE}=N\textsubscript{0}\Big(\frac{dN}{dE}\Big)\bigg|_{file}
\end{equation}

\begin{table*}
	\centering

	\begin{tabular}{c c }
        \hline
        Energy & Differential Flux\\
        MeV & cm\textsuperscript{-2} s\textsuperscript{-1} MeV\textsuperscript{-1}\\
	    \hline
50& 2 $\times$ 10 \textsuperscript{-3}\\
60& 8 $\times$ 10 \textsuperscript{-4}\\
70& 4 $\times$ 10 \textsuperscript{-4}\\
80& 1 $\times$ 10 \textsuperscript{-4}\\
90& 6 $\times$ 10 \textsuperscript{-5}\\
100& 2 $\times$ 10 \textsuperscript{-5}\\
200& 1 $\times$ 10 \textsuperscript{-11}\\
		\hline
	\end{tabular}
    	\caption{Definition of the FileFunction spectral model with differential flux at a given energy }
        \label{tab:FileFunction}
\end{table*}

\subsection{The Effect of Pulsar Core Temperature}
\label{sec:discussion_core_temperature}

The emission rate for axions is strongly dependent on pulsar core temperature, \textit{T\textsubscript{c}}, being proportional to  \textit{T\textsubscript{c}\textsuperscript{6}} [\onlinecite{RN213}]. We therefore re-examine the applicable value of \textit{T\textsubscript{c}} for modeling axion emission and the effect of lowering \textit{T\textsubscript{c}} on that emission. The authors of [\onlinecite{RN131}] select \textit{T\textsubscript{c}}=20 MeV on the basis of the range temperatures applicable to equation of state (EOS)  simulations of pulsar degenerate matter [\onlinecite{RN261,RN291,RN292}], slower neutron star cooling due to super-fluidity [\onlinecite{RN262,RN294}] and surface temperature observations of the pulsar J0953+0755 [\onlinecite{RN295}].  

We now consider to what extent the works cited above explicitly support the choice of \textit{T\textsubscript{c}}=20 MeV. In EOS modeling both [\onlinecite{RN261}] and [\onlinecite{RN291}] use \textit{T\textsubscript{c}} as a free model parameter (in the range 0$-$60 MeV and 0$-$15 MeV respectively) for the construction of phase diagrams but this does not indicate a preferential value for \textit{T\textsubscript{c}}. In [\onlinecite{RN292}], a specific Fermi temperature of \textit{T\textsubscript{F}} of 20 MeV per nucleon is supported but no preferred value of \textit{T\textsubscript{c}} is indicated. The cooling of quark hybrid (QH) stars (a special case of a higher density neutron star where quarks experience deconfinement from nucleons) is considered in [\onlinecite{RN262}] with QH stars in fact cooling \textit{more} quickly than hadron neutron stars unless a colour flavour locked (CFL) quark phase with a higher CFL gap parameter of 1 MeV is considered. However, by 10\textsuperscript{5} yr all modelled QH stars again exhibit \textit{greater} cooling then hadron neutron stars. As all neutron stars in our pulsar sample have age \textgreater 10\textsuperscript{5} yr (Table~\ref{tab:psr_list}), this QH star slow cooling regime will not result in a higher value for \textit{T\textsubscript{c}} in our sample than might be expected from normal cooling processes.  The discussion of crustal heating arising from super fluidity in neutron stars also refutes \textit{T\textsubscript{c}}=20 MeV,  with one neutron star J0953+0755 (PSR 0950+08) analysed in [\onlinecite{RN131}] having an internal temperature of between 0.09 keV  and 0.11 keV [\onlinecite{RN294}]. Although there is more recent evidence of internal heating of J0953+0755 from far UV HST observations (surface temperature (ST) = (1$-$3) $\times$ 10\textsuperscript{5} K [\onlinecite{RN256}] vs 7 $\times$ 10\textsuperscript{4} K of [\onlinecite{RN295}]), this would still only result in a maximum \textit{T\textsubscript{c}} of 1.34 keV assuming \textit{T\textsubscript{c}}=12 $\times$ (ST/10\textsuperscript{6} K)\textsuperscript{1.82} keV [\onlinecite{RN294,RN301}]. 

\textcolor{black}{The authors of [\onlinecite{RN344}] have modelled the cooling of neutron stars using a fully general relativistic stellar evolution code, without exotic cooling, allowing for inputs for equations of state and uncertainties in superfluidity along with a finite time scale of thermal conduction. They determine \textit{T\textsubscript{c}} to be initially 3.98 $\times$ 10\textsuperscript{9} K (343 keV) when the neutron star is 9 hours old, decreasing to  1.99 $\times$ 10\textsuperscript{9} K (171 keV) at 1 yr, 6.31 $\times$ 10\textsuperscript{8} K (54 keV) at 1000 yr  and 1.99 $\times$ 10\textsuperscript{8} K (17 keV)) at 10\textsuperscript{5} yr. This cooling trend agrees well with the modelling of pulsar cooling in [\onlinecite{PSR_REVIEW_YAKOVLEV}] where the highest pulsar surface temperatures (in all scenarios) of 3.98 $\times$ 10\textsuperscript{6} K at 1 yr and 1.99 $\times$ 10\textsuperscript{6} K at 10\textsuperscript{5} yr yield a \textit{T\textsubscript{c}} of 148 keV and 12 keV respectively using the ST to \textit{T\textsubscript{c}} conversion above.  It should also be noted that \textit{Chandra} observations of the very young pulsar Cas A (age $\approx$ 330 yr), yield an ST of  2.04 $\times$ 10\textsuperscript{6} K [\onlinecite{RN343}] equivalent to \textit{T\textsubscript{c}} = 43.9 keV using the ST to \textit{T\textsubscript{c}} conversion above. Similarly, in their modeling of Cas A cooling using the observations of [\onlinecite{RN343}], the author of [\onlinecite{RN342}] determines the \textit{T\textsubscript{c}} of Cas A to be 7.2 $\times$ 10\textsuperscript{8} K, equivalent to 62 keV. }

We therefore consider \textit{T\textsubscript{c}}=20 MeV to be a high temperature choice more consistent with the neutron star core just after the supernova event. In [\onlinecite{RN297}], EOS and hydrodynamic modeling is performed in the first second after the supernova core bounce and proto neutron star (PNS) creation. Here, at 150 ms post bounce, \textit{T\textsubscript{c}} can be 14 MeV at the core, falling to 10 MeV at a radius of 10 km, before rising to a peak of 32 MeV at radius 12 km. Other modeling work demonstrates that a peak PNS  \textit{T\textsubscript{c}} of 30 to 43 MeV is possible, falling to 5 to 18 MeV  within 50 s [\onlinecite{RN296}] due to efficient cooling by neutrino emission. A very short time later, at 120 s, the PNS \textit{T\textsubscript{c}} is 2.2 MeV [\onlinecite{Nakazato2018}]. This suggests that plausible values of \textit{T\textsubscript{c}} are much less than 20 MeV with \textit{T\textsubscript{c}}=1 MeV being achieved within seconds [\onlinecite{RN265}].

\begin{figure*}
	\includegraphics[width=12cm,keepaspectratio]{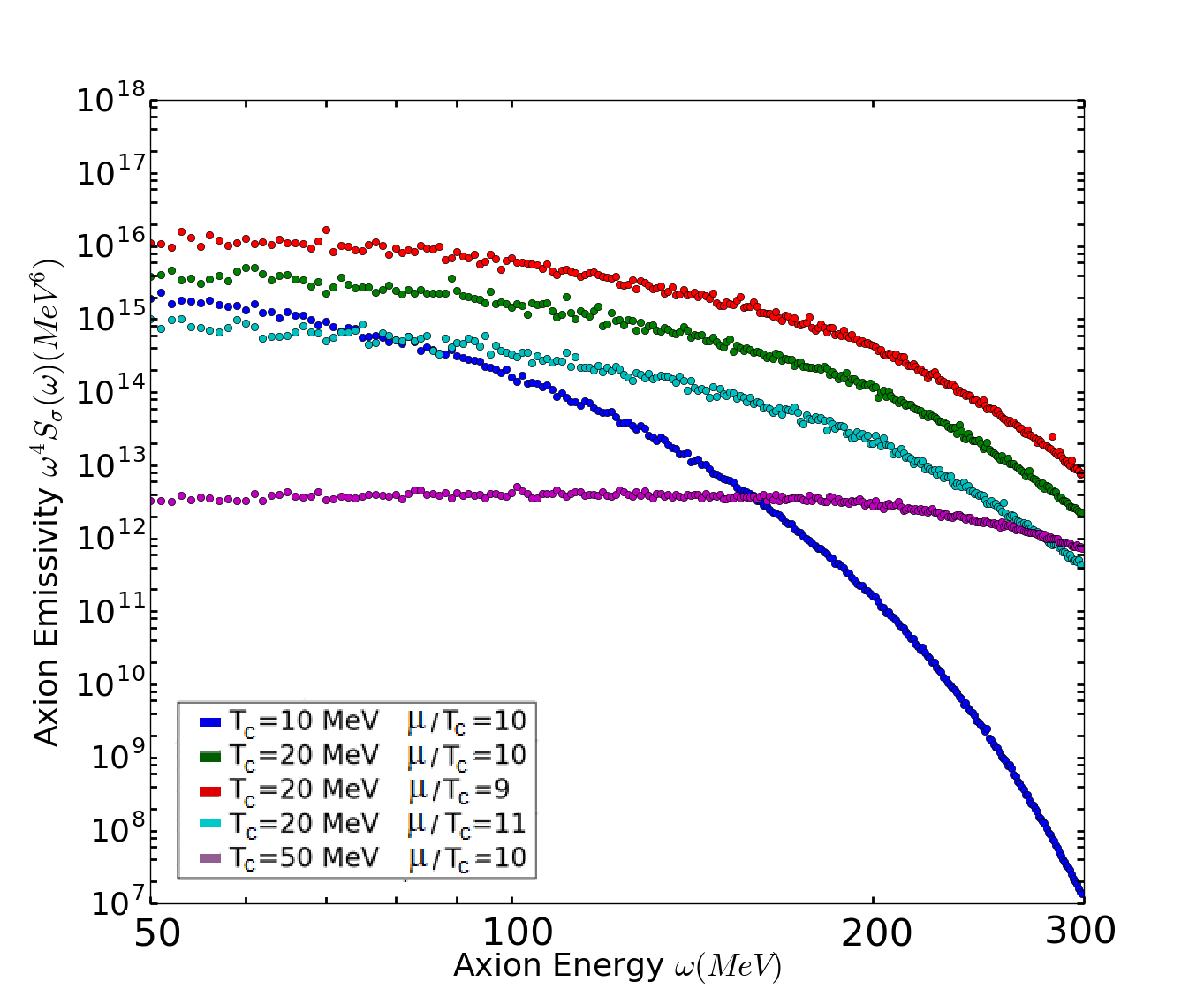}
    \caption{The energy dependence of axion emissivity \textit{$\omega$\textsuperscript{4}S\textsubscript{$\sigma$}($\omega$)} on axion energy \textit{$\omega$} for varying pulsar core temperature \textit{T\textsubscript{c}} and $\mu$/\textit{T\textsubscript{c}} derived by Monte Carlo numerical integration of an analytic simplification of S\textsubscript{$\sigma$}($\omega$).}
    \label{fig:SpinStructurePlot}
\end{figure*}

We re-evaluate \textit{$\omega$\textsuperscript{4}S\textsubscript{$\sigma$}($\omega$)}, on which the axion emissivity depends (Eqn.~\ref{AxionEmissivity}), for \textit{T\textsubscript{c}} \textless 20 MeV. We use the analytic simplification for the phase space integral for \textit{S\textsubscript{$\sigma$}($\omega$)} from [\onlinecite{RN281}] and perform a 5 dimensional numeric Monte Carlo integration as described in the Appendix \ref{appendix:AppendixA}. In order to check our method we first reproduce the  \textit{$\omega$\textsuperscript{4}S\textsubscript{$\sigma$}($\omega$)} plot from [\onlinecite{RN131}]  using a \textit{T\textsubscript{c}} of 10$-$50 MeV, $\mu$/\textit{T\textsubscript{c}} = 9$-$11 and \textit{p\textsubscript{Fn}} = 300 MeV (Fig.~\ref{fig:SpinStructurePlot}). 

We reproduce the essential features of the [\onlinecite{RN131}] plot both in magnitude and in the following respects: 
\begin{itemize}
  \item Increasing the value of $\mu$/\textit{T\textsubscript{c}} for fixed \textit{T\textsubscript{c}}=20 MeV decreases amplitude of \textit{$\omega$\textsuperscript{4}S\textsubscript{$\sigma$}($\omega$)}
  \item \textit{$\omega$\textsuperscript{4}S\textsubscript{$\sigma$}($\omega$)} for \textit{T\textsubscript{c}}=10 MeV cuts-off at a lower value of $\omega$=100 MeV than for \textit{T\textsubscript{c}}=20 MeV
  \item The \textit{T\textsubscript{c}}=50 MeV case has lower values of \textit{$\omega$\textsuperscript{4}S\textsubscript{$\sigma$}($\omega$)} than the \textit{T\textsubscript{c}}=20 MeV case, with \textit{$\omega$\textsuperscript{4}S\textsubscript{$\sigma$}($\omega$)} remaining broadly flat across higher $\omega$ values of 100$-$300 MeV with no pronounced cut-off at 200$-$300 MeV
  \item The value of \textit{$\omega$\textsuperscript{4}S\textsubscript{$\sigma$}($\omega$)} spans one order of magnitude for the 20 MeV case and varying $\mu$/\textit{T\textsubscript{c}} = 9$-$11
\end{itemize}

\begin{figure*}
	\includegraphics[width=12cm,keepaspectratio]{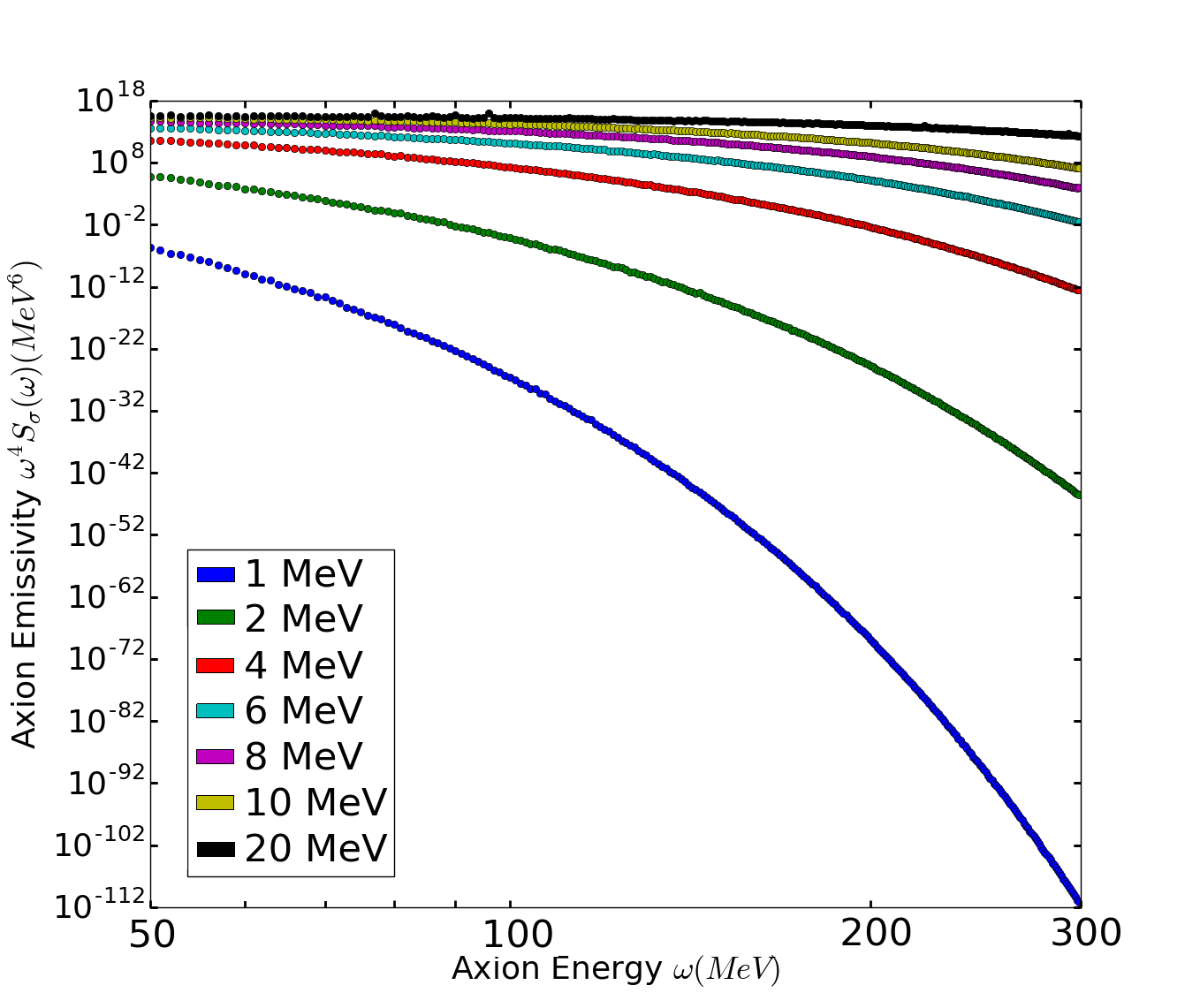}
    \caption{The energy dependence of axion emissivity \textit{$\omega$\textsuperscript{4}S\textsubscript{$\sigma$}($\omega$)} on axion energy \textit{$\omega$} for \textit{T\textsubscript{c}} = 1-20 MeV and $\mu$/\textit{T\textsubscript{c}} =10 derived by Monte Carlo numerical integration of an analytic simplification of S\textsubscript{$\sigma$}($\omega$). Reducing \textit{T\textsubscript{c}} from 20 MeV to 4 MeV lowers emissivity by a factor of 10\textsuperscript{8} at \textit{$\omega$}=100 MeV.}
    \label{fig:SpinStructurePlotLT}
\end{figure*}

We then evaluate \textit{$\omega$\textsuperscript{4}S\textsubscript{$\sigma$}($\omega$)}, in a lower temperature regime, for \textit{p\textsubscript{Fn}} = 300 MeV, $\mu$/\textit{T\textsubscript{c}} = 10 and consider lower pulsar core temperatures with \textit{T\textsubscript{c}} = 1$-$20 MeV (Fig.~\ref{fig:SpinStructurePlotLT}). Lowering \textit{T\textsubscript{c}} from 20 MeV to a plausible PNS temperature of 4 MeV reduces axion emissivity and hence gamma-ray emission by a factor of 10\textsuperscript{8} for axions of energy $\omega$=100 MeV. It therefore seems implausible that there would be detectable gamma-ray emission to allow the determination of \textit{m\textsubscript{a}} using the astrophysical model of gamma-ray emission from [\onlinecite{RN131}] (Eqn.~\ref{PhotonFlux}), for realistic pulsar core temperatures. We note however that this model is based on a quite conservative assumption that gamma-ray emission arises solely from axion radiative decay as opposed to axion to gamma-ray photon conversion in the \textit{B} field of the pulsar. It is therefore possible that an alternative model allowing axion to photon conversion could produce detectable gamma-ray emission.

\begin{figure*}
	\includegraphics[width=20cm,keepaspectratio]{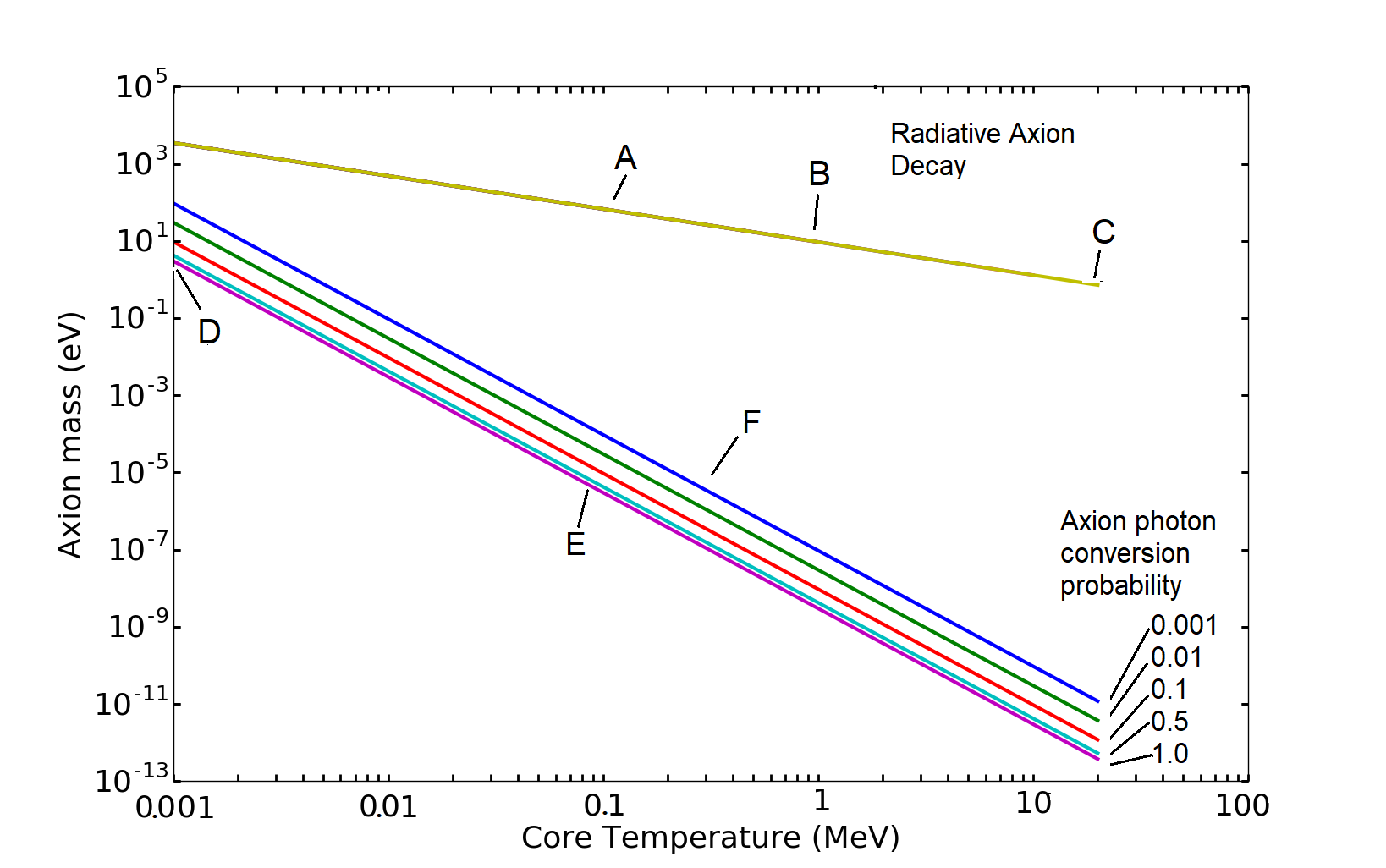}
    \caption{Plot of axion mass with respect to \textit{T\textsubscript{c}} using an alternative energy loss rate model and varying axion to photon conversion probabilities from 0.001 to 1.00. Also shown is the more conservative axion radiative decay case (top). \textcolor{black}{At realistic values of \textit{T\textsubscript{c}} of 0.1 and 1 MeV, radiative decay alone yields unrealistic values for UL \textit{m\textsubscript{a}} of 67.5 eV and 9.4 eV respectively (labelled A and B). At an unrealistic high  value of 20 MeV for \textit{T\textsubscript{c}} the UL \textit{m\textsubscript{a}} is 0.7 eV (Labelled C) . At \textit{T\textsubscript{c}}=1 keV, UL \textit{m\textsubscript{a}} is 3.0 eV, assuming total axion to photon conversion (labelled D). To keep UL \textit{m\textsubscript{a}}\textgreater 10\textsuperscript{-6} eV, which is the classic axion search lower bound, requires \textit{T\textsubscript{c}} \textless 0.1 MeV (labelled E) or \textit{T\textsubscript{c}} \textless 0.4 MeV with a low axion to photon conversion probability of 0.001 (labelled F).  }     }
    \label{fig:RaffeltConversionPlot}
\end{figure*}

The probable lack of detectable gamma-ray emission in the lower temperature regime leads us to derive values for UL \textit{m\textsubscript{a}} from an alternative model (Eqn.~\ref{UL_Axion_Mass}) based on the axion power equation which defines an energy loss rate due to axion production in the pulsar core (Eqn.~\ref{Raffelt}). Using the UL gamma-ray luminosity (Table~\ref{tab:psr_results}) we determine UL \textit{m\textsubscript{a}} from Eqn.~\ref{UL_Axion_Mass} whilst varying \textit{T\textsubscript{c}} and the probability of axion to photon conversion in the pulsar \textit{B} field. On Fig.~\ref{fig:RaffeltConversionPlot} we show the range of UL \textit{m\textsubscript{a}} values that we obtain. \textcolor{black}{We see that the conversion of axions to gamma-ray photons via radiative decay results in the highest UL \textit{m\textsubscript{a}} (67.5 eV at 0.1 MeV, 9.4 eV at 1 MeV and 0.7 eV at 20 MeV, points A, B and C respectively) which is above the classic  \textit{m\textsubscript{a}} search range of 10\textsuperscript{-2}$-$10\textsuperscript{-6} eV. Similarly by varying the axion to photon conversion probability from 0.001 to 1.0 (total conversion), we only obtain an UL \textit{m\textsubscript{a}} above the lower search bound of 10\textsuperscript{-6} eV for \textit{T\textsubscript{c}} \textless 0.1 MeV independent of the degree of axion to photon conversion or \textit{T\textsubscript{c}} \textless 0.4 MeV assuming a probability of $\leq$ 0.001 for axion to photon conversion (Points E and F of Fig.~\ref{fig:RaffeltConversionPlot} respectively). At \textit{T\textsubscript{c}}=1 keV the lowest UL \textit{m\textsubscript{a}} obtainable would be 3.0 eV assuming total conversion of axions to photons (Point D of Fig.~\ref{fig:RaffeltConversionPlot}). We do not offer a view on the degree of axion to photon conversion in the pulsar B field but simply present a range of conversion alternatives to give indicative values of the UL \textit{m\textsubscript{a}}. }

The determination of a plausible and precise UL \textit{m\textsubscript{a}} from this alternative model thus requires both realistic lower values of \textit{T\textsubscript{c}} and a knowledge of the precise extent of the axion to photon conversion in the pulsar \textit{B} field. We have dealt with the value of \textit{T\textsubscript{c}} in the PNS and old pulsar cases above; however, whilst [\onlinecite{RN131}] consider there to be no axion to photon conversion in the pulsar \textit{B} field (using vacuum bi-refringence arguments) there is no consensus on the extent of axion to gamma-ray photon conversion in pulsar \textit{B} fields. More attention has been paid to axion to X-ray photon inter-conversion in pulsars [\onlinecite{RN227}] and in axion like particle (ALP)  to X-ray conversion in the higher \textit{B} field (20 $\times$ 10\textsuperscript{14} G) of magnetars by [\onlinecite{RN299}]. [\onlinecite{RN299}] finds \textit{P\textsubscript{$a\rightarrow \gamma$}}=0.225 for $\omega$ = 3 keV (the peak emission) and \textit{P\textsubscript{$a\rightarrow \gamma$}} = 0.025 for $\omega$ = 200 keV when \textit{T\textsubscript{c}}=50$-$250 keV. The lower \textit{B} field of our sample notwithstanding (average $B$=2.78 $\times$ 10\textsuperscript{12} G) such values of \textit{P\textsubscript{$a\rightarrow \gamma$}} and \textit{T\textsubscript{c}} could yield constraints on  \textit{m\textsubscript{a}} in the classic axion search range using the alternative model (Fig.~\ref{fig:RaffeltConversionPlot}). 

Finally, the normalized axion energy spectrum \textit{dN\textsubscript{a}/d$\omega$} peaks at $\omega$/\textit{T\textsubscript{c}} = 2 [\onlinecite{RAFFELT1996}]. This implies that the photon energy spectrum would peak at energy \textit{T\textsubscript{c}}. Therefore for the values of \textit{T\textsubscript{c}} discussed above, in the 1 MeV range or below, the determination of an UL for  unpulsed gamma-ray emission in  \textcolor{black}{our pulsar sample or preferably younger pulsars with a potentially higher \textit{T\textsubscript{c}}}, by future low-energy gamma-ray observatories such as the All-Sky Medium Energy Gamma-ray observatory (AMEGO) or e-ASTROGAM, with greater sensitivity then any current observatory in the 0.2$-$10 MeV band [\onlinecite{AMEGO}, \onlinecite{DEANGELIS20181}] \textcolor{black}{may} allow an improved determination on the UL \textit{m\textsubscript{a}} presented in this work.

\begin{table*}
	\centering
	\begin{tabular}{c c c c c}
        \hline
        Pulsar  & UL Photon Flux (60-200 MeV)  & UL Photon Flux (60-500 MeV) & UL \textit{m\textsubscript{a}}  $\omega$=100 MeV & UL \textit{m\textsubscript{a}}  $\omega$=200 MeV\\
         & (From [\onlinecite{RN131}])  & This analysis  & (10\textsuperscript{-2} eV)  & (10\textsuperscript{-2} eV)\\
        
                & (10\textsuperscript{-9} cm\textsuperscript{-2} s\textsuperscript{-1}) & (10\textsuperscript{-9} cm\textsuperscript{-2} s\textsuperscript{-1}) & & \\

        \hline

           J0108-1431	&	4.03	&	1.75  & 0.69 & 2.31\\
           J0953+0755	&	7.40	&	4.75  & 0.97 & 3.26\\
           J0630-2834	&	4.82	&	18.90 & 0.97 & 3.25\\
           J1136+1551	&	8.52	&	5.01  & 1.25 & 4.17\\

	\hline
	\end{tabular}
    	
\caption{The UL photon flux for 4 pulsars from [\onlinecite{RN131}] (60-200 MeV) compared to our analysis (60-500 MeV) and UL \textit{m\textsubscript{a}} which we derive from [\onlinecite{RN131}] fluxes for axions of energy 100 MeV and 200 MeV using Eqn.~\ref{AxionMassFromPhotonFlux}.}
        \label{tab:berenji_fluxes}
\end{table*}

\section{Conclusions}
\label{sec:Conclusion}

We analyze data from 17 nearby pulsars using 9 years of \textit{Fermi}-LAT data and detect none. Using the UL photon flux and the astrophysical model of [\onlinecite{RN131}] which assumes a pulsar core temperature of 20 MeV we determine an improved UL axion mass (\textit{m\textsubscript{a}}) of 0.96 and 3.21 $\times$ 10\textsuperscript{-2} eV for axions of energy 100 MeV and 200 MeV respectively. However, we show that at realistic pulsar core temperatures of \textless 4 MeV, axion emissivity is so reduced that is unlikely a reasonable determination of UL \textit{m\textsubscript{a}} can be made with this method. An alternative axion energy loss rate model yields a plausible range of UL \textit{m\textsubscript{a}} values assuming low pulsar core temperatures but requires both the core temperature and the axion to photon conversion probability to be known to set a useful limit. Observation of the un-pulsed gamma-ray emission of our selected pulsar sample with future medium energy gamma-ray observatories such as AMEGO and e-ASTROGAM \textcolor{black}{may} allow a better determination of UL \textit{m\textsubscript{a}}.

\section*{Acknowledgements}

We acknowledge the excellent data and analysis tools provided by the \textit{Fermi}-LAT collaboration. AMB and PMC acknowledge the financial support of the UK Science and Technology Facilities Council consolidated grant ST/P000541/1. This research has made use of the SIMBAD database, operated at CDS, Strasbourg, France (\onlinecite{2000A&AS..143....9W}). \textcolor{black}{Finally we thank the anonymous referee for their review and very useful comments which improved this paper.}

\appendix*
\section{NUCLEON PHASE SPACE INTEGRATION}
\label{appendix:AppendixA}

The spin structure function of  Eqn.~\ref{SpinStructure} has an analytic simplification as presented by [\onlinecite{RN281}] of which we repeat the relevant points here. From the original 12-dimensional integral, 7 dimensions may be integrated out analytically so that a 5-dimensional integral remains to be solved through numerical integration (as opposed to numerical integration of the 4-dimensional integral of [\onlinecite{RN281}]). 
  
Firstly the 3-dimensional momentum delta function is used to integrate out $d^3\bm{p_{4}}$. Then, the non-relativistic nucleons have energy $E_i=p_i^2/2M_N$ and so the energy balance term c

\begin{equation*}
\begin{aligned}
E_{1}+E_{2}-E_{3}-E_{4}+\omega
\end{aligned}
\end{equation*}

\begin{equation}{\label{A1}}
    \begin{aligned}
    =\frac{-2p_{3}^2-2\mathbf{p_{1}}\cdot\mathbf{p_{2}}+2\mathbf{p_{1}}\cdot\mathbf{p_{3}}+2\mathbf{p_{2}}\cdot\mathbf{p_{3}}} {2M_N} +\omega
    \end{aligned} 
\end{equation}

Next, a polar co-ordinate system is used with $\alpha$ and $\beta$ being the polar and azimuthmal angles of $\bm{p_{2}}$ relative to $\bm{p_{1}}$  and $\theta$ and $\Phi$ those of $\bm{p_{3}}$. The medium is isotropic so the $\bm{p_{1}}$ momentum can be chosen in the $z$ direction so $\int d^3\bm{p_{1}}=4\pi\int dp_1$ with $p_1=|\bm{p_{1}}| $. The medium isototropy also allows the azimuthmal angle $d\Phi$ to be trivially integrated to leave three nontrivial angular integrations with the remaining angular variables expressed as follows:

\begin{equation}{\label{A4}}
    \begin{aligned}
    \mathbf{p_{1}}\cdot\mathbf{p_{2}}\:=\: p_1p_2\:cos\:\alpha
    \end{aligned} 
\end{equation}

\begin{equation}{\label{A5}}
    \begin{aligned}
    \mathbf{p_{1}}\cdot\mathbf{p_{3}}\:=\: p_1p_3\:cos\:\theta
    \end{aligned} 
\end{equation}

\begin{equation}{\label{A6}}
    \begin{aligned}
    \mathbf{p_{2}}\cdot\mathbf{p_{3}}\:=\: p_2p_3\:cos\:\alpha\:cos\:\theta+\:sin\:\alpha+\:sin\:\theta+\:cos\:\beta
    \end{aligned} 
\end{equation}

The integration over $d\beta$ is carried out using the $\delta$ function with $f(\beta)\equiv E_1 + E_2 - E_3 - E_4 + \omega$  and $\beta_1$ being the root of $f(\beta)=0$ in the interval [0,$\pi$] giving:

\begin{equation}{\label{A8}}
    \begin{aligned}
    \int_{0}^{2\pi} \: d\beta \: \delta [f(\beta)]\:=\:   \frac{2}{|df(\beta)/d\beta|_{\beta=\beta_1}} \:  \Theta \Big (\Big{|}\frac{d f(\beta)}{d\beta} \Big {|}_{\beta=\beta_1}^2\Big)
    \end{aligned} 
\end{equation}

The derivative can be expressed as

\begin{equation}{\label{A9}}
    \begin{aligned}
    \Big{|}\frac{d f(\beta)}{d\beta} \Big {|}_{\beta=\beta_1}= \sqrt{az^2+bz+c}
    \end{aligned} 
\end{equation}

where 

\begin{equation}
    \begin{aligned}
     z \equiv cos \: \alpha \\
    \end{aligned} 
\end{equation}

\begin{equation}
    \begin{aligned}
     a=p_2^2(-p_1^2-p_3^2+2p_1p_3cos\theta) \\
    \end{aligned} 
\end{equation}

\begin{equation}
    \begin{aligned}
     b=2\omega M_Np_1p_2-2p_1p_2p_3^2-2\omega M_Np_2p_3cos\theta \\ +2p_1^2p_3cos\theta+2p_2p_3^3cos\theta-2p_1p_2p_3^2cos^2\theta \\
    \end{aligned} 
\end{equation}

\begin{equation}
    \begin{aligned}
     c=\omega^2M_N^2+2\omega M_Np_3^2+p_2^2p_3^2-p_3^4-2\omega M_Np_1p_3cos\theta \\ +2p_1p_3^3cos \theta-p_1^2p_3^2cos^2\theta-p_2^2p_3^3cos^2\theta\\
    \end{aligned} 
\end{equation}

Finally the analytic simplification of equation~\ref{A8} can be solved by numerical integration through a Monte Carlo method integrating over $d p_1d p_2d p_3 d \: cos \: \theta \: d \: cos \: \alpha$.

\clearpage
\bibliography{exportlist.bib}

\begin{thebibliography}{68}%
\makeatletter
\providecommand \@ifxundefined [1]{%
 \@ifx{#1\undefined}
}%
\providecommand \@ifnum [1]{%
 \ifnum #1\expandafter \@firstoftwo
 \else \expandafter \@secondoftwo
 \fi
}%
\providecommand \@ifx [1]{%
 \ifx #1\expandafter \@firstoftwo
 \else \expandafter \@secondoftwo
 \fi
}%
\providecommand \natexlab [1]{#1}%
\providecommand \enquote  [1]{``#1''}%
\providecommand \bibnamefont  [1]{#1}%
\providecommand \bibfnamefont [1]{#1}%
\providecommand \citenamefont [1]{#1}%
\providecommand \href@noop [0]{\@secondoftwo}%
\providecommand \href [0]{\begingroup \@sanitize@url \@href}%
\providecommand \@href[1]{\@@startlink{#1}\@@href}%
\providecommand \@@href[1]{\endgroup#1\@@endlink}%
\providecommand \@sanitize@url [0]{\catcode `\\12\catcode `\$12\catcode
  `\&12\catcode `\#12\catcode `\^12\catcode `\_12\catcode `\%12\relax}%
\providecommand \@@startlink[1]{}%
\providecommand \@@endlink[0]{}%
\providecommand \url  [0]{\begingroup\@sanitize@url \@url }%
\providecommand \@url [1]{\endgroup\@href {#1}{\urlprefix }}%
\providecommand \urlprefix  [0]{URL }%
\providecommand \Eprint [0]{\href }%
\providecommand \doibase [0]{https://doi.org/}%
\providecommand \selectlanguage [0]{\@gobble}%
\providecommand \bibinfo  [0]{\@secondoftwo}%
\providecommand \bibfield  [0]{\@secondoftwo}%
\providecommand \translation [1]{[#1]}%
\providecommand \BibitemOpen [0]{}%
\providecommand \bibitemStop [0]{}%
\providecommand \bibitemNoStop [0]{.\EOS\space}%
\providecommand \EOS [0]{\spacefactor3000\relax}%
\providecommand \BibitemShut  [1]{\csname bibitem#1\endcsname}%
\let\auto@bib@innerbib\@empty
\bibitem [{\citenamefont {Peccei}\ and\ \citenamefont
  {Quinn}(1977)}]{PhysRevLett.38.1440}%
  \BibitemOpen
  \bibfield  {author} {\bibinfo {author} {\bibfnamefont {R.~D.}\ \bibnamefont
  {Peccei}}and\ \bibinfo {author} {\bibfnamefont {H.~R.}\ \bibnamefont
  {Quinn}},\ }\bibfield  {title} {\enquote {\bibinfo {title} {$\mathrm{CP}$
  conservation in the presence of pseudoparticles},}\ }\href
  {https://doi.org/10.1103/PhysRevLett.38.1440} {\bibfield  {journal} {\bibinfo
   {journal} {Phys. Rev. Lett.}\ }\textbf {\bibinfo {volume} {38}},\ \bibinfo
  {pages} {1440--1443} (\bibinfo {year} {1977})}\BibitemShut {NoStop}%
\bibitem [{\citenamefont {Weinberg}(1978)}]{PhysRevLett.40.223}%
  \BibitemOpen
  \bibfield  {author} {\bibinfo {author} {\bibfnamefont {S.}~\bibnamefont
  {Weinberg}},\ }\bibfield  {title} {\enquote {\bibinfo {title} {A new light
  boson?}}\ }\href {https://doi.org/10.1103/PhysRevLett.40.223} {\bibfield
  {journal} {\bibinfo  {journal} {Phys. Rev. Lett.}\ }\textbf {\bibinfo
  {volume} {40}},\ \bibinfo {pages} {223--226} (\bibinfo {year}
  {1978})}\BibitemShut {NoStop}%
\bibitem [{\citenamefont {Dine}, \citenamefont {Fischler},\ and\ \citenamefont
  {Srednicki}(1981)}]{Dine:1981rt}%
  \BibitemOpen
  \bibfield  {author} {\bibinfo {author} {\bibfnamefont {M.}~\bibnamefont
  {Dine}}, \bibinfo {author} {\bibfnamefont {W.}~\bibnamefont {Fischler}}, and\
  \bibinfo {author} {\bibfnamefont {M.}~\bibnamefont {Srednicki}},\ }\bibfield
  {title} {\enquote {\bibinfo {title} {{A Simple Solution to the Strong CP
  Problem with a Harmless Axion}},}\ }\href
  {https://doi.org/10.1016/0370-2693(81)90590-6} {\bibfield  {journal}
  {\bibinfo  {journal} {Phys. Lett.}\ }\textbf {\bibinfo {volume} {104B}},\
  \bibinfo {pages} {199--202} (\bibinfo {year} {1981})}\BibitemShut {NoStop}%
\bibitem [{\citenamefont {Keil}\ \emph {et~al.}(1997)\citenamefont {Keil},
  \citenamefont {Janka}, \citenamefont {Schramm}, \citenamefont {Sigl},
  \citenamefont {Turner},\ and\ \citenamefont {Ellis}}]{PhysRevD.56.2419}%
  \BibitemOpen
  \bibfield  {author} {\bibinfo {author} {\bibfnamefont {W.}~\bibnamefont
  {Keil}}, \bibinfo {author} {\bibfnamefont {H.-T.}\ \bibnamefont {Janka}},
  \bibinfo {author} {\bibfnamefont {D.~N.}\ \bibnamefont {Schramm}}, \bibinfo
  {author} {\bibfnamefont {G.}~\bibnamefont {Sigl}}, \bibinfo {author}
  {\bibfnamefont {M.~S.}\ \bibnamefont {Turner}}, and\ \bibinfo {author}
  {\bibfnamefont {J.}~\bibnamefont {Ellis}},\ }\bibfield  {title} {\enquote
  {\bibinfo {title} {Fresh look at axions and sn 1987a},}\ }\href
  {https://doi.org/10.1103/PhysRevD.56.2419} {\bibfield  {journal} {\bibinfo
  {journal} {Phys. Rev. D}\ }\textbf {\bibinfo {volume} {56}},\ \bibinfo
  {pages} {2419--2432} (\bibinfo {year} {1997})}\BibitemShut {NoStop}%
\bibitem [{\citenamefont {Rybka}(2014)}]{RYBKA201414}%
  \BibitemOpen
  \bibfield  {author} {\bibinfo {author} {\bibfnamefont {G.}~\bibnamefont
  {Rybka}},\ }\bibfield  {title} {\enquote {\bibinfo {title} {Direct detection
  searches for axion dark matter},}\ }\href
  {https://doi.org/https://doi.org/10.1016/j.dark.2014.05.003} {\bibfield
  {journal} {\bibinfo  {journal} {Physics of the Dark Universe}\ }\textbf
  {\bibinfo {volume} {4}},\ \bibinfo {pages} {14 -- 16} (\bibinfo {year}
  {2014})},\ \bibinfo {note} {dARK TAUP2013}\BibitemShut {NoStop}%
\bibitem [{\citenamefont {Asztalos}\ \emph {et~al.}(2010)\citenamefont
  {Asztalos}, \citenamefont {Carosi}, \citenamefont {Hagmann}, \citenamefont
  {Kinion}, \citenamefont {van Bibber}, \citenamefont {Hotz}, \citenamefont
  {Rosenberg}, \citenamefont {Rybka}, \citenamefont {Hoskins}, \citenamefont
  {Hwang}, \citenamefont {Sikivie}, \citenamefont {Tanner}, \citenamefont
  {Bradley},\ and\ \citenamefont {Clarke}}]{PhysRevLett.104.041301}%
  \BibitemOpen
  \bibfield  {author} {\bibinfo {author} {\bibfnamefont {S.~J.}\ \bibnamefont
  {Asztalos}}, \bibinfo {author} {\bibfnamefont {G.}~\bibnamefont {Carosi}},
  \bibinfo {author} {\bibfnamefont {C.}~\bibnamefont {Hagmann}}, \bibinfo
  {author} {\bibfnamefont {D.}~\bibnamefont {Kinion}}, \bibinfo {author}
  {\bibfnamefont {K.}~\bibnamefont {van Bibber}}, \bibinfo {author}
  {\bibfnamefont {M.}~\bibnamefont {Hotz}}, \bibinfo {author} {\bibfnamefont
  {L.~J.}\ \bibnamefont {Rosenberg}}, \bibinfo {author} {\bibfnamefont
  {G.}~\bibnamefont {Rybka}}, \bibinfo {author} {\bibfnamefont
  {J.}~\bibnamefont {Hoskins}}, \bibinfo {author} {\bibfnamefont
  {J.}~\bibnamefont {Hwang}}, \bibinfo {author} {\bibfnamefont
  {P.}~\bibnamefont {Sikivie}}, \bibinfo {author} {\bibfnamefont {D.~B.}\
  \bibnamefont {Tanner}}, \bibinfo {author} {\bibfnamefont {R.}~\bibnamefont
  {Bradley}}, and\ \bibinfo {author} {\bibfnamefont {J.}~\bibnamefont
  {Clarke}},\ }\bibfield  {title} {\enquote {\bibinfo {title} {Squid-based
  microwave cavity search for dark-matter axions},}\ }\href
  {https://doi.org/10.1103/PhysRevLett.104.041301} {\bibfield  {journal}
  {\bibinfo  {journal} {Phys. Rev. Lett.}\ }\textbf {\bibinfo {volume} {104}},\
  \bibinfo {pages} {041301} (\bibinfo {year} {2010})}\BibitemShut {NoStop}%
\bibitem [{\citenamefont {Duffy}\ \emph {et~al.}(2006)\citenamefont {Duffy},
  \citenamefont {Sikivie}, \citenamefont {Tanner}, \citenamefont {Asztalos},
  \citenamefont {Hagmann}, \citenamefont {Kinion}, \citenamefont {Rosenberg},
  \citenamefont {van Bibber}, \citenamefont {Yu},\ and\ \citenamefont
  {Bradley}}]{PhysRevD.74.012006}%
  \BibitemOpen
  \bibfield  {author} {\bibinfo {author} {\bibfnamefont {L.~D.}\ \bibnamefont
  {Duffy}}, \bibinfo {author} {\bibfnamefont {P.}~\bibnamefont {Sikivie}},
  \bibinfo {author} {\bibfnamefont {D.~B.}\ \bibnamefont {Tanner}}, \bibinfo
  {author} {\bibfnamefont {S.~J.}\ \bibnamefont {Asztalos}}, \bibinfo {author}
  {\bibfnamefont {C.}~\bibnamefont {Hagmann}}, \bibinfo {author} {\bibfnamefont
  {D.}~\bibnamefont {Kinion}}, \bibinfo {author} {\bibfnamefont {L.~J.}\
  \bibnamefont {Rosenberg}}, \bibinfo {author} {\bibfnamefont {K.}~\bibnamefont
  {van Bibber}}, \bibinfo {author} {\bibfnamefont {D.~B.}\ \bibnamefont {Yu}},
  and\ \bibinfo {author} {\bibfnamefont {R.~F.}\ \bibnamefont {Bradley}},\
  }\bibfield  {title} {\enquote {\bibinfo {title} {High resolution search for
  dark-matter axions},}\ }\href {https://doi.org/10.1103/PhysRevD.74.012006}
  {\bibfield  {journal} {\bibinfo  {journal} {Phys. Rev. D}\ }\textbf {\bibinfo
  {volume} {74}},\ \bibinfo {pages} {012006} (\bibinfo {year}
  {2006})}\BibitemShut {NoStop}%
\bibitem [{\citenamefont {Hagmann}\ \emph {et~al.}(1998)\citenamefont
  {Hagmann}, \citenamefont {Kinion}, \citenamefont {Stoeffl}, \citenamefont
  {van Bibber}, \citenamefont {Daw}, \citenamefont {Peng}, \citenamefont
  {Rosenberg}, \citenamefont {LaVeigne}, \citenamefont {Sikivie}, \citenamefont
  {Sullivan}, \citenamefont {Tanner}, \citenamefont {Nezrick}, \citenamefont
  {Turner}, \citenamefont {Moltz}, \citenamefont {Powell},\ and\ \citenamefont
  {Golubev}}]{PhysRevLett.80.2043}%
  \BibitemOpen
  \bibfield  {author} {\bibinfo {author} {\bibfnamefont {C.}~\bibnamefont
  {Hagmann}}, \bibinfo {author} {\bibfnamefont {D.}~\bibnamefont {Kinion}},
  \bibinfo {author} {\bibfnamefont {W.}~\bibnamefont {Stoeffl}}, \bibinfo
  {author} {\bibfnamefont {K.}~\bibnamefont {van Bibber}}, \bibinfo {author}
  {\bibfnamefont {E.}~\bibnamefont {Daw}}, \bibinfo {author} {\bibfnamefont
  {H.}~\bibnamefont {Peng}}, \bibinfo {author} {\bibfnamefont {L.~J.}\
  \bibnamefont {Rosenberg}}, \bibinfo {author} {\bibfnamefont {J.}~\bibnamefont
  {LaVeigne}}, \bibinfo {author} {\bibfnamefont {P.}~\bibnamefont {Sikivie}},
  \bibinfo {author} {\bibfnamefont {N.~S.}\ \bibnamefont {Sullivan}}, \bibinfo
  {author} {\bibfnamefont {D.~B.}\ \bibnamefont {Tanner}}, \bibinfo {author}
  {\bibfnamefont {F.}~\bibnamefont {Nezrick}}, \bibinfo {author} {\bibfnamefont
  {M.~S.}\ \bibnamefont {Turner}}, \bibinfo {author} {\bibfnamefont {D.~M.}\
  \bibnamefont {Moltz}}, \bibinfo {author} {\bibfnamefont {J.}~\bibnamefont
  {Powell}}, and\ \bibinfo {author} {\bibfnamefont {N.~A.}\ \bibnamefont
  {Golubev}},\ }\bibfield  {title} {\enquote {\bibinfo {title} {Results from a
  high-sensitivity search for cosmic axions},}\ }\href
  {https://doi.org/10.1103/PhysRevLett.80.2043} {\bibfield  {journal} {\bibinfo
   {journal} {Phys. Rev. Lett.}\ }\textbf {\bibinfo {volume} {80}},\ \bibinfo
  {pages} {2043--2046} (\bibinfo {year} {1998})}\BibitemShut {NoStop}%
\bibitem [{\citenamefont {Hoskins}\ \emph {et~al.}(2011)\citenamefont
  {Hoskins}, \citenamefont {Hwang}, \citenamefont {Martin}, \citenamefont
  {Sikivie}, \citenamefont {Sullivan}, \citenamefont {Tanner}, \citenamefont
  {Hotz}, \citenamefont {Rosenberg}, \citenamefont {Rybka}, \citenamefont
  {Wagner}, \citenamefont {Asztalos}, \citenamefont {Carosi}, \citenamefont
  {Hagmann}, \citenamefont {Kinion}, \citenamefont {van Bibber}, \citenamefont
  {Bradley},\ and\ \citenamefont {Clarke}}]{PhysRevD.84.121302}%
  \BibitemOpen
  \bibfield  {author} {\bibinfo {author} {\bibfnamefont {J.}~\bibnamefont
  {Hoskins}}, \bibinfo {author} {\bibfnamefont {J.}~\bibnamefont {Hwang}},
  \bibinfo {author} {\bibfnamefont {C.}~\bibnamefont {Martin}}, \bibinfo
  {author} {\bibfnamefont {P.}~\bibnamefont {Sikivie}}, \bibinfo {author}
  {\bibfnamefont {N.~S.}\ \bibnamefont {Sullivan}}, \bibinfo {author}
  {\bibfnamefont {D.~B.}\ \bibnamefont {Tanner}}, \bibinfo {author}
  {\bibfnamefont {M.}~\bibnamefont {Hotz}}, \bibinfo {author} {\bibfnamefont
  {L.~J.}\ \bibnamefont {Rosenberg}}, \bibinfo {author} {\bibfnamefont
  {G.}~\bibnamefont {Rybka}}, \bibinfo {author} {\bibfnamefont
  {A.}~\bibnamefont {Wagner}}, \bibinfo {author} {\bibfnamefont {S.~J.}\
  \bibnamefont {Asztalos}}, \bibinfo {author} {\bibfnamefont {G.}~\bibnamefont
  {Carosi}}, \bibinfo {author} {\bibfnamefont {C.}~\bibnamefont {Hagmann}},
  \bibinfo {author} {\bibfnamefont {D.}~\bibnamefont {Kinion}}, \bibinfo
  {author} {\bibfnamefont {K.}~\bibnamefont {van Bibber}}, \bibinfo {author}
  {\bibfnamefont {R.}~\bibnamefont {Bradley}}, and\ \bibinfo {author}
  {\bibfnamefont {J.}~\bibnamefont {Clarke}},\ }\bibfield  {title} {\enquote
  {\bibinfo {title} {Search for nonvirialized axionic dark matter},}\ }\href
  {https://doi.org/10.1103/PhysRevD.84.121302} {\bibfield  {journal} {\bibinfo
  {journal} {Phys. Rev. D}\ }\textbf {\bibinfo {volume} {84}},\ \bibinfo
  {pages} {121302} (\bibinfo {year} {2011})}\BibitemShut {NoStop}%
\bibitem [{\citenamefont {Du}\ \emph {et~al.}(2018)\citenamefont {Du},
  \citenamefont {Force}, \citenamefont {Khatiwada}, \citenamefont {Lentz},
  \citenamefont {Ottens}, \citenamefont {Rosenberg}, \citenamefont {Rybka},
  \citenamefont {Carosi}, \citenamefont {Woollett}, \citenamefont {Bowring},
  \citenamefont {Chou}, \citenamefont {Sonnenschein}, \citenamefont {Wester},
  \citenamefont {Boutan}, \citenamefont {Oblath}, \citenamefont {Bradley},
  \citenamefont {Daw}, \citenamefont {Dixit}, \citenamefont {Clarke},
  \citenamefont {O'Kelley}, \citenamefont {Crisosto}, \citenamefont {Gleason},
  \citenamefont {Jois}, \citenamefont {Sikivie}, \citenamefont {Stern},
  \citenamefont {Sullivan}, \citenamefont {Tanner},\ and\ \citenamefont
  {Hilton}}]{ADMX}%
  \BibitemOpen
  \bibfield  {author} {\bibinfo {author} {\bibfnamefont {N.}~\bibnamefont
  {Du}}, \bibinfo {author} {\bibfnamefont {N.}~\bibnamefont {Force}}, \bibinfo
  {author} {\bibfnamefont {R.}~\bibnamefont {Khatiwada}}, \bibinfo {author}
  {\bibfnamefont {E.}~\bibnamefont {Lentz}}, \bibinfo {author} {\bibfnamefont
  {R.}~\bibnamefont {Ottens}}, \bibinfo {author} {\bibfnamefont {L.~J.}\
  \bibnamefont {Rosenberg}}, \bibinfo {author} {\bibfnamefont {G.}~\bibnamefont
  {Rybka}}, \bibinfo {author} {\bibfnamefont {G.}~\bibnamefont {Carosi}},
  \bibinfo {author} {\bibfnamefont {N.}~\bibnamefont {Woollett}}, \bibinfo
  {author} {\bibfnamefont {D.}~\bibnamefont {Bowring}}, \bibinfo {author}
  {\bibfnamefont {A.~S.}\ \bibnamefont {Chou}}, \bibinfo {author}
  {\bibfnamefont {A.}~\bibnamefont {Sonnenschein}}, \bibinfo {author}
  {\bibfnamefont {W.}~\bibnamefont {Wester}}, \bibinfo {author} {\bibfnamefont
  {C.}~\bibnamefont {Boutan}}, \bibinfo {author} {\bibfnamefont {N.~S.}\
  \bibnamefont {Oblath}}, \bibinfo {author} {\bibfnamefont {R.}~\bibnamefont
  {Bradley}}, \bibinfo {author} {\bibfnamefont {E.~J.}\ \bibnamefont {Daw}},
  \bibinfo {author} {\bibfnamefont {A.~V.}\ \bibnamefont {Dixit}}, \bibinfo
  {author} {\bibfnamefont {J.}~\bibnamefont {Clarke}}, \bibinfo {author}
  {\bibfnamefont {S.~R.}\ \bibnamefont {O'Kelley}}, \bibinfo {author}
  {\bibfnamefont {N.}~\bibnamefont {Crisosto}}, \bibinfo {author}
  {\bibfnamefont {J.~R.}\ \bibnamefont {Gleason}}, \bibinfo {author}
  {\bibfnamefont {S.}~\bibnamefont {Jois}}, \bibinfo {author} {\bibfnamefont
  {P.}~\bibnamefont {Sikivie}}, \bibinfo {author} {\bibfnamefont
  {I.}~\bibnamefont {Stern}}, \bibinfo {author} {\bibfnamefont {N.~S.}\
  \bibnamefont {Sullivan}}, \bibinfo {author} {\bibfnamefont {D.~B.}\
  \bibnamefont {Tanner}}, and\ \bibinfo {author} {\bibfnamefont {G.~C.}\
  \bibnamefont {Hilton}} (\bibinfo {collaboration} {ADMX Collaboration}),\
  }\bibfield  {title} {\enquote {\bibinfo {title} {Search for invisible axion
  dark matter with the axion dark matter experiment},}\ }\href
  {https://doi.org/10.1103/PhysRevLett.120.151301} {\bibfield  {journal}
  {\bibinfo  {journal} {Phys. Rev. Lett.}\ }\textbf {\bibinfo {volume} {120}},\
  \bibinfo {pages} {151301} (\bibinfo {year} {2018})}\BibitemShut {NoStop}%
\bibitem [{\citenamefont {Sedrakian}(2016)}]{PhysRevD.93.065044}%
  \BibitemOpen
  \bibfield  {author} {\bibinfo {author} {\bibfnamefont {A.}~\bibnamefont
  {Sedrakian}},\ }\bibfield  {title} {\enquote {\bibinfo {title} {Axion cooling
  of neutron stars},}\ }\href {https://doi.org/10.1103/PhysRevD.93.065044}
  {\bibfield  {journal} {\bibinfo  {journal} {Phys. Rev. D}\ }\textbf {\bibinfo
  {volume} {93}},\ \bibinfo {pages} {065044} (\bibinfo {year}
  {2016})}\BibitemShut {NoStop}%
\bibitem [{\citenamefont {Berenji}, \citenamefont {Gaskins},\ and\
  \citenamefont {Meyer}(2016)}]{RN131}%
  \BibitemOpen
  \bibfield  {author} {\bibinfo {author} {\bibfnamefont {B.}~\bibnamefont
  {Berenji}}, \bibinfo {author} {\bibfnamefont {J.}~\bibnamefont {Gaskins}},
  and\ \bibinfo {author} {\bibfnamefont {M.}~\bibnamefont {Meyer}},\ }\bibfield
   {title} {\enquote {\bibinfo {title} {Constraints on axions and axionlike
  particles from {Fermi} large area telescope observations of neutron stars},}\
  }\href {https://doi.org/10.1103/PhysRevD.93.045019} {\bibfield  {journal}
  {\bibinfo  {journal} {Physical Review D}\ }\textbf {\bibinfo {volume} {93}},\
  \bibinfo {pages} {13} (\bibinfo {year} {2016})}\BibitemShut {NoStop}%
\bibitem [{\citenamefont {Iwamoto}(1984)}]{RN290}%
  \BibitemOpen
  \bibfield  {author} {\bibinfo {author} {\bibfnamefont {N.}~\bibnamefont
  {Iwamoto}},\ }\bibfield  {title} {\enquote {\bibinfo {title} {Axion emission
  from neutron stars},}\ }\href {https://doi.org/10.1103/PhysRevLett.53.1198}
  {\bibfield  {journal} {\bibinfo  {journal} {Physical Review Letters}\
  }\textbf {\bibinfo {volume} {53}},\ \bibinfo {pages} {1198--1201} (\bibinfo
  {year} {1984})}\BibitemShut {NoStop}%
\bibitem [{\citenamefont {Iwamoto}(2001)}]{RN219}%
  \BibitemOpen
  \bibfield  {author} {\bibinfo {author} {\bibfnamefont {N.}~\bibnamefont
  {Iwamoto}},\ }\bibfield  {title} {\enquote {\bibinfo {title} {Nucleon-nucleon
  bremsstrahlung of axions and pseudoscalar particles from neutron-star
  matter},}\ }\href {https://doi.org/10.1103/PhysRevD.64.043002} {\bibfield
  {journal} {\bibinfo  {journal} {Physical Review D}\ }\textbf {\bibinfo
  {volume} {64}},\ \bibinfo {pages} {10} (\bibinfo {year} {2001})}\BibitemShut
  {NoStop}%
\bibitem [{\citenamefont {Hanhart}, \citenamefont {Phillips},\ and\
  \citenamefont {Reddy}(2001)}]{RN260}%
  \BibitemOpen
  \bibfield  {author} {\bibinfo {author} {\bibfnamefont {C.}~\bibnamefont
  {Hanhart}}, \bibinfo {author} {\bibfnamefont {D.~R.}\ \bibnamefont
  {Phillips}}, and\ \bibinfo {author} {\bibfnamefont {S.}~\bibnamefont
  {Reddy}},\ }\bibfield  {title} {\enquote {\bibinfo {title} {Neutrino and
  axion emissivities of neutron stars from nucleon-nucleon scattering data},}\
  }\href {https://doi.org/10.1016/s0370-2693(00)01382-4} {\bibfield  {journal}
  {\bibinfo  {journal} {Physics Letters B}\ }\textbf {\bibinfo {volume}
  {499}},\ \bibinfo {pages} {9--15} (\bibinfo {year} {2001})}\BibitemShut
  {NoStop}%
\bibitem [{\citenamefont {Hannestad}\ and\ \citenamefont
  {Raffelt}(1998)}]{RN281}%
  \BibitemOpen
  \bibfield  {author} {\bibinfo {author} {\bibfnamefont {S.}~\bibnamefont
  {Hannestad}}and\ \bibinfo {author} {\bibfnamefont {G.}~\bibnamefont
  {Raffelt}},\ }\bibfield  {title} {\enquote {\bibinfo {title} {Supernova
  neutrino opacity from nucleon-nucleon bremsstrahlung and related
  processes},}\ }\href {https://doi.org/10.1086/306303} {\bibfield  {journal}
  {\bibinfo  {journal} {Astrophysical Journal}\ }\textbf {\bibinfo {volume}
  {507}},\ \bibinfo {pages} {339--352} (\bibinfo {year} {1998})}\BibitemShut
  {NoStop}%
\bibitem [{\citenamefont {Brinkmann}\ and\ \citenamefont
  {Turner}(1988)}]{RN213}%
  \BibitemOpen
  \bibfield  {author} {\bibinfo {author} {\bibfnamefont {R.~P.}\ \bibnamefont
  {Brinkmann}}and\ \bibinfo {author} {\bibfnamefont {M.~S.}\ \bibnamefont
  {Turner}},\ }\bibfield  {title} {\enquote {\bibinfo {title} {Numerical rates
  for nucleon-nucleon, axion bremsstrahlung},}\ }\href
  {https://doi.org/10.1103/PhysRevD.38.2338} {\bibfield  {journal} {\bibinfo
  {journal} {Physical Review D}\ }\textbf {\bibinfo {volume} {38}},\ \bibinfo
  {pages} {2338--2348} (\bibinfo {year} {1988})}\BibitemShut {NoStop}%
\bibitem [{\citenamefont {Mayle}\ \emph {et~al.}(1988)\citenamefont {Mayle},
  \citenamefont {Wilson}, \citenamefont {Ellis}, \citenamefont {Olive},
  \citenamefont {Schramm},\ and\ \citenamefont {Steigman}}]{MAYLE1988188}%
  \BibitemOpen
  \bibfield  {author} {\bibinfo {author} {\bibfnamefont {R.}~\bibnamefont
  {Mayle}}, \bibinfo {author} {\bibfnamefont {J.~R.}\ \bibnamefont {Wilson}},
  \bibinfo {author} {\bibfnamefont {J.}~\bibnamefont {Ellis}}, \bibinfo
  {author} {\bibfnamefont {K.}~\bibnamefont {Olive}}, \bibinfo {author}
  {\bibfnamefont {D.~N.}\ \bibnamefont {Schramm}}, and\ \bibinfo {author}
  {\bibfnamefont {G.}~\bibnamefont {Steigman}},\ }\bibfield  {title} {\enquote
  {\bibinfo {title} {Constraints on axions from sn 1987a},}\ }\href
  {https://doi.org/https://doi.org/10.1016/0370-2693(88)91595-X} {\bibfield
  {journal} {\bibinfo  {journal} {Physics Letters B}\ }\textbf {\bibinfo
  {volume} {203}},\ \bibinfo {pages} {188 -- 196} (\bibinfo {year}
  {1988})}\BibitemShut {NoStop}%
\bibitem [{\citenamefont {G.~Raffelt}(1996)}]{RAFFELT1996}%
  \BibitemOpen
  \bibfield  {author} {\bibinfo {author} {\bibfnamefont {G.}~\bibnamefont
  {G.~Raffelt}},\ }\bibfield  {title} {\enquote {\bibinfo {title} {Stars as
  laboratories for fundamental physics: The astrophysics of neutrinos, axions,
  and other weakly interacting particles},}\ }\href@noop {} {\bibfield
  {journal} {\bibinfo  {journal} {Bibliovault OAI Repository, the University of
  Chicago Press}\ } (\bibinfo {year} {1996})}\BibitemShut {NoStop}%
\bibitem [{\citenamefont {Raffelt}(2008)}]{RN341}%
  \BibitemOpen
  \bibfield  {author} {\bibinfo {author} {\bibfnamefont {G.~G.}\ \bibnamefont
  {Raffelt}},\ }\bibfield  {title} {\enquote {\bibinfo {title} {Astrophysical
  axion bounds},}\ }\href {https://doi.org/10.1007/978-3-540-73518-2-3}
  {\bibfield  {journal} {\bibinfo  {journal} {Axions: Theory, Cosmology, and
  Experimental Searches}\ }\textbf {\bibinfo {volume} {741}},\ \bibinfo {pages}
  {51--71} (\bibinfo {year} {2008})}\BibitemShut {NoStop}%
\bibitem [{\citenamefont {{Manchester}}\ \emph {et~al.}(2005)\citenamefont
  {{Manchester}}, \citenamefont {{Hobbs}}, \citenamefont {{Teoh}},\ and\
  \citenamefont {{Hobbs}}}]{ATNF1}%
  \BibitemOpen
  \bibfield  {author} {\bibinfo {author} {\bibfnamefont {R.~N.}\ \bibnamefont
  {{Manchester}}}, \bibinfo {author} {\bibfnamefont {G.~B.}\ \bibnamefont
  {{Hobbs}}}, \bibinfo {author} {\bibfnamefont {A.}~\bibnamefont {{Teoh}}},
  and\ \bibinfo {author} {\bibfnamefont {M.}~\bibnamefont {{Hobbs}}},\
  }\bibfield  {title} {\enquote {\bibinfo {title} {{The Australia Telescope
  National Facility Pulsar Catalogue}},}\ }\href
  {https://doi.org/10.1086/428488} {\bibfield  {journal} {\bibinfo  {journal}
  {The Astronomical Journal}\ }\textbf {\bibinfo {volume} {129}},\ \bibinfo
  {pages} {1993--2006} (\bibinfo {year} {2005})},\ \Eprint
  {https://arxiv.org/abs/astro-ph/0412641} {astro-ph/0412641} \BibitemShut
  {NoStop}%
\bibitem [{Note1()}]{Note1}%
  \BibitemOpen
  \bibinfo {note}
  {Http://www.atnf.csiro.au/research/pulsar/psrcat/}\BibitemShut {NoStop}%
\bibitem [{Note2()}]{Note2}%
  \BibitemOpen
  \bibinfo {note} {Https://confluence.slac.stanford.edu/display/GLAMCOG/ \\
  Public+List+of+LAT-Detected+Gamma-Ray+Pulsars, list last updated 19\protect
  \textsuperscript {th} Oct 2018, accessed on 14\protect \textsuperscript {th}
  Feb 2019}\BibitemShut {NoStop}%
\bibitem [{\citenamefont {Abdo}\ \emph {et~al.}(2013)\citenamefont {Abdo},
  \citenamefont {Ajello}, \citenamefont {Allafort}, \citenamefont {Baldini},
  \citenamefont {Ballet}, \citenamefont {Barbiellini}, \citenamefont {Baring},
  \citenamefont {Bastieri}, \citenamefont {Belfiore}, \citenamefont
  {Bellazzini}, \citenamefont {Bhattacharyya}, \citenamefont {Bissaldi},
  \citenamefont {Bloom}, \citenamefont {Bonamente} \emph {et~al.}}]{RN244}%
  \BibitemOpen
  \bibfield  {author} {\bibinfo {author} {\bibfnamefont {A.~A.}\ \bibnamefont
  {Abdo}}, \bibinfo {author} {\bibfnamefont {M.}~\bibnamefont {Ajello}},
  \bibinfo {author} {\bibfnamefont {A.}~\bibnamefont {Allafort}}, \bibinfo
  {author} {\bibfnamefont {L.}~\bibnamefont {Baldini}}, \bibinfo {author}
  {\bibfnamefont {J.}~\bibnamefont {Ballet}}, \bibinfo {author} {\bibfnamefont
  {G.}~\bibnamefont {Barbiellini}}, \bibinfo {author} {\bibfnamefont {M.~G.}\
  \bibnamefont {Baring}}, \bibinfo {author} {\bibfnamefont {D.}~\bibnamefont
  {Bastieri}}, \bibinfo {author} {\bibfnamefont {A.}~\bibnamefont {Belfiore}},
  \bibinfo {author} {\bibfnamefont {R.}~\bibnamefont {Bellazzini}}, \bibinfo
  {author} {\bibfnamefont {B.}~\bibnamefont {Bhattacharyya}}, \bibinfo {author}
  {\bibfnamefont {E.}~\bibnamefont {Bissaldi}}, \bibinfo {author}
  {\bibfnamefont {E.~D.}\ \bibnamefont {Bloom}}, \bibinfo {author}
  {\bibfnamefont {E.}~\bibnamefont {Bonamente}},  \emph {et~al.},\ }\bibfield
  {title} {\enquote {\bibinfo {title} {The second {Fermi} large area telescope
  catalog of gamma-ray pulsars},}\ }\href
  {https://doi.org/10.1088/0067-0049/208/2/17} {\bibfield  {journal} {\bibinfo
  {journal} {Astrophysical Journal Supplement Series}\ }\textbf {\bibinfo
  {volume} {208}},\ \bibinfo {pages} {59} (\bibinfo {year} {2013})}\BibitemShut
  {NoStop}%
\bibitem [{\citenamefont {Burke-Spolaor}\ and\ \citenamefont
  {Bailes}(2010)}]{bb10}%
  \BibitemOpen
  \bibfield  {author} {\bibinfo {author} {\bibfnamefont {S.}~\bibnamefont
  {Burke-Spolaor}}and\ \bibinfo {author} {\bibfnamefont {M.}~\bibnamefont
  {Bailes}},\ }\bibfield  {title} {\enquote {\bibinfo {title} {The millisecond
  radio sky: transients from a blind single-pulse search},}\ }\href
  {https://doi.org/10.1111/j.1365-2966.2009.15965.x} {\bibfield  {journal}
  {\bibinfo  {journal} {Monthly Notices of the Royal Astronomical Society}\
  }\textbf {\bibinfo {volume} {402}},\ \bibinfo {pages} {855--866} (\bibinfo
  {year} {2010})}\BibitemShut {NoStop}%
\bibitem [{\citenamefont {Jiang}\ \emph {et~al.}(2017)\citenamefont {Jiang},
  \citenamefont {Cui}, \citenamefont {Schmid}, \citenamefont {McLaughlin},\
  and\ \citenamefont {Cao}}]{jcs+17}%
  \BibitemOpen
  \bibfield  {author} {\bibinfo {author} {\bibfnamefont {M.}~\bibnamefont
  {Jiang}}, \bibinfo {author} {\bibfnamefont {B.~Y.}\ \bibnamefont {Cui}},
  \bibinfo {author} {\bibfnamefont {N.~A.}\ \bibnamefont {Schmid}}, \bibinfo
  {author} {\bibfnamefont {M.~A.}\ \bibnamefont {McLaughlin}}, and\ \bibinfo
  {author} {\bibfnamefont {Z.~C.}\ \bibnamefont {Cao}},\ }\bibfield  {title}
  {\enquote {\bibinfo {title} {Wavelet denoising of radio observations of
  rotating radio transients (rrats): Improved timing parameters for eight
  rrats},}\ }\href {https://doi.org/10.3847/1538-4357/aa88c3} {\bibfield
  {journal} {\bibinfo  {journal} {Astrophysical Journal}\ }\textbf {\bibinfo
  {volume} {847}},\ \bibinfo {pages} {13} (\bibinfo {year} {2017})}\BibitemShut
  {NoStop}%
\bibitem [{\citenamefont {Bailes}\ \emph {et~al.}(1997)\citenamefont {Bailes},
  \citenamefont {Johnston}, \citenamefont {Bell}, \citenamefont {Lorimer},
  \citenamefont {Stappers}, \citenamefont {Manchester}, \citenamefont {Lyne},
  \citenamefont {Nicastro}, \citenamefont {Damico},\ and\ \citenamefont
  {Gaensler}}]{bjb+97}%
  \BibitemOpen
  \bibfield  {author} {\bibinfo {author} {\bibfnamefont {M.}~\bibnamefont
  {Bailes}}, \bibinfo {author} {\bibfnamefont {S.}~\bibnamefont {Johnston}},
  \bibinfo {author} {\bibfnamefont {J.~F.}\ \bibnamefont {Bell}}, \bibinfo
  {author} {\bibfnamefont {D.~R.}\ \bibnamefont {Lorimer}}, \bibinfo {author}
  {\bibfnamefont {B.~W.}\ \bibnamefont {Stappers}}, \bibinfo {author}
  {\bibfnamefont {R.~N.}\ \bibnamefont {Manchester}}, \bibinfo {author}
  {\bibfnamefont {A.~G.}\ \bibnamefont {Lyne}}, \bibinfo {author}
  {\bibfnamefont {L.}~\bibnamefont {Nicastro}}, \bibinfo {author}
  {\bibfnamefont {N.}~\bibnamefont {Damico}}, and\ \bibinfo {author}
  {\bibfnamefont {B.~M.}\ \bibnamefont {Gaensler}},\ }\bibfield  {title}
  {\enquote {\bibinfo {title} {Discovery of four isolated millisecond
  pulsars},}\ }\href {https://doi.org/10.1086/304041} {\bibfield  {journal}
  {\bibinfo  {journal} {Astrophysical Journal}\ }\textbf {\bibinfo {volume}
  {481}},\ \bibinfo {pages} {386--391} (\bibinfo {year} {1997})}\BibitemShut
  {NoStop}%
\bibitem [{\citenamefont {Reardon}\ \emph {et~al.}(2016)\citenamefont
  {Reardon}, \citenamefont {Hobbs}, \citenamefont {Coles}, \citenamefont
  {Levin}, \citenamefont {Keith}, \citenamefont {Bailes}, \citenamefont {Bhat},
  \citenamefont {Burke-Spolaor}, \citenamefont {Dai}, \citenamefont {Kerr},
  \citenamefont {Lasky}, \citenamefont {Manchester}, \citenamefont {Oslowski},
  \citenamefont {Ravi}, \citenamefont {Shannon}, \citenamefont {van Straten},
  \citenamefont {Toomey}, \citenamefont {Wang}, \citenamefont {Wen},
  \citenamefont {You},\ and\ \citenamefont {Zhu}}]{rhc+16}%
  \BibitemOpen
  \bibfield  {author} {\bibinfo {author} {\bibfnamefont {D.~J.}\ \bibnamefont
  {Reardon}}, \bibinfo {author} {\bibfnamefont {G.}~\bibnamefont {Hobbs}},
  \bibinfo {author} {\bibfnamefont {W.}~\bibnamefont {Coles}}, \bibinfo
  {author} {\bibfnamefont {Y.}~\bibnamefont {Levin}}, \bibinfo {author}
  {\bibfnamefont {M.~J.}\ \bibnamefont {Keith}}, \bibinfo {author}
  {\bibfnamefont {M.}~\bibnamefont {Bailes}}, \bibinfo {author} {\bibfnamefont
  {N.~D.~R.}\ \bibnamefont {Bhat}}, \bibinfo {author} {\bibfnamefont
  {S.}~\bibnamefont {Burke-Spolaor}}, \bibinfo {author} {\bibfnamefont
  {S.}~\bibnamefont {Dai}}, \bibinfo {author} {\bibfnamefont {M.}~\bibnamefont
  {Kerr}}, \bibinfo {author} {\bibfnamefont {P.~D.}\ \bibnamefont {Lasky}},
  \bibinfo {author} {\bibfnamefont {R.~N.}\ \bibnamefont {Manchester}},
  \bibinfo {author} {\bibfnamefont {S.}~\bibnamefont {Oslowski}}, \bibinfo
  {author} {\bibfnamefont {V.}~\bibnamefont {Ravi}}, \bibinfo {author}
  {\bibfnamefont {R.~M.}\ \bibnamefont {Shannon}}, \bibinfo {author}
  {\bibfnamefont {W.}~\bibnamefont {van Straten}}, \bibinfo {author}
  {\bibfnamefont {L.}~\bibnamefont {Toomey}}, \bibinfo {author} {\bibfnamefont
  {J.}~\bibnamefont {Wang}}, \bibinfo {author} {\bibfnamefont {L.}~\bibnamefont
  {Wen}}, \bibinfo {author} {\bibfnamefont {X.~P.}\ \bibnamefont {You}}, and\
  \bibinfo {author} {\bibfnamefont {X.~J.}\ \bibnamefont {Zhu}},\ }\bibfield
  {title} {\enquote {\bibinfo {title} {Timing analysis for 20 millisecond
  pulsars in the {Parkes} pulsar timing array},}\ }\href
  {https://doi.org/10.1093/mnras/stv2395} {\bibfield  {journal} {\bibinfo
  {journal} {Monthly Notices of the Royal Astronomical Society}\ }\textbf
  {\bibinfo {volume} {455}},\ \bibinfo {pages} {1751--1769} (\bibinfo {year}
  {2016})}\BibitemShut {NoStop}%
\bibitem [{\citenamefont {Manchester}\ \emph {et~al.}(1978)\citenamefont
  {Manchester}, \citenamefont {Lyne}, \citenamefont {Taylor}, \citenamefont
  {Durdin}, \citenamefont {Large},\ and\ \citenamefont {Little}}]{mlt+78}%
  \BibitemOpen
  \bibfield  {author} {\bibinfo {author} {\bibfnamefont {R.~N.}\ \bibnamefont
  {Manchester}}, \bibinfo {author} {\bibfnamefont {A.~G.}\ \bibnamefont
  {Lyne}}, \bibinfo {author} {\bibfnamefont {J.~H.}\ \bibnamefont {Taylor}},
  \bibinfo {author} {\bibfnamefont {J.~M.}\ \bibnamefont {Durdin}}, \bibinfo
  {author} {\bibfnamefont {M.~I.}\ \bibnamefont {Large}}, and\ \bibinfo
  {author} {\bibfnamefont {A.~G.}\ \bibnamefont {Little}},\ }\bibfield  {title}
  {\enquote {\bibinfo {title} {2nd {Molonglo} pulsar survey - discovery of 155
  pulsars},}\ }\href {https://doi.org/10.1093/mnras/185.2.409} {\bibfield
  {journal} {\bibinfo  {journal} {Monthly Notices of the Royal Astronomical
  Society}\ }\textbf {\bibinfo {volume} {185}},\ \bibinfo {pages} {409--421}
  (\bibinfo {year} {1978})}\BibitemShut {NoStop}%
\bibitem [{\citenamefont {Siegman}, \citenamefont {Manchester},\ and\
  \citenamefont {Durdin}(1993)}]{smd93}%
  \BibitemOpen
  \bibfield  {author} {\bibinfo {author} {\bibfnamefont {B.~C.}\ \bibnamefont
  {Siegman}}, \bibinfo {author} {\bibfnamefont {R.~N.}\ \bibnamefont
  {Manchester}}, and\ \bibinfo {author} {\bibfnamefont {J.~M.}\ \bibnamefont
  {Durdin}},\ }\bibfield  {title} {\enquote {\bibinfo {title} {Timing
  parameters for 59 pulsars},}\ }\href
  {https://doi.org/10.1093/mnras/262.2.449} {\bibfield  {journal} {\bibinfo
  {journal} {Monthly Notices of the Royal Astronomical Society}\ }\textbf
  {\bibinfo {volume} {262}},\ \bibinfo {pages} {449--455} (\bibinfo {year}
  {1993})}\BibitemShut {NoStop}%
\bibitem [{\citenamefont {Manchester}\ \emph {et~al.}(1996)\citenamefont
  {Manchester}, \citenamefont {Lyne}, \citenamefont {Damico}, \citenamefont
  {Bailes}, \citenamefont {Johnston}, \citenamefont {Lorimer}, \citenamefont
  {Harrison}, \citenamefont {Nicastro},\ and\ \citenamefont {Bell}}]{mld+96}%
  \BibitemOpen
  \bibfield  {author} {\bibinfo {author} {\bibfnamefont {R.~N.}\ \bibnamefont
  {Manchester}}, \bibinfo {author} {\bibfnamefont {A.~G.}\ \bibnamefont
  {Lyne}}, \bibinfo {author} {\bibfnamefont {N.}~\bibnamefont {Damico}},
  \bibinfo {author} {\bibfnamefont {M.}~\bibnamefont {Bailes}}, \bibinfo
  {author} {\bibfnamefont {S.}~\bibnamefont {Johnston}}, \bibinfo {author}
  {\bibfnamefont {D.~R.}\ \bibnamefont {Lorimer}}, \bibinfo {author}
  {\bibfnamefont {P.~A.}\ \bibnamefont {Harrison}}, \bibinfo {author}
  {\bibfnamefont {L.}~\bibnamefont {Nicastro}}, and\ \bibinfo {author}
  {\bibfnamefont {J.~F.}\ \bibnamefont {Bell}},\ }\bibfield  {title} {\enquote
  {\bibinfo {title} {The {Parkes} southern pulsar survey .1. observing and data
  analysis systems and initial results},}\ }\href
  {https://doi.org/10.1093/mnras/279.4.1235} {\bibfield  {journal} {\bibinfo
  {journal} {Monthly Notices of the Royal Astronomical Society}\ }\textbf
  {\bibinfo {volume} {279}},\ \bibinfo {pages} {1235--1250} (\bibinfo {year}
  {1996})}\BibitemShut {NoStop}%
\bibitem [{\citenamefont {Hobbs}\ \emph {et~al.}(2004)\citenamefont {Hobbs},
  \citenamefont {Lyne}, \citenamefont {Kramer}, \citenamefont {Martin},\ and\
  \citenamefont {Jordan}}]{hlk+04}%
  \BibitemOpen
  \bibfield  {author} {\bibinfo {author} {\bibfnamefont {G.}~\bibnamefont
  {Hobbs}}, \bibinfo {author} {\bibfnamefont {A.~G.}\ \bibnamefont {Lyne}},
  \bibinfo {author} {\bibfnamefont {M.}~\bibnamefont {Kramer}}, \bibinfo
  {author} {\bibfnamefont {C.~E.}\ \bibnamefont {Martin}}, and\ \bibinfo
  {author} {\bibfnamefont {C.}~\bibnamefont {Jordan}},\ }\bibfield  {title}
  {\enquote {\bibinfo {title} {Long-term timing observations of 374 pulsars},}\
  }\href {https://doi.org/10.1111/j.1365-2966.2004.08157.x} {\bibfield
  {journal} {\bibinfo  {journal} {Monthly Notices of the Royal Astronomical
  Society}\ }\textbf {\bibinfo {volume} {353}},\ \bibinfo {pages} {1311--1344}
  (\bibinfo {year} {2004})}\BibitemShut {NoStop}%
\bibitem [{\citenamefont {Pilkington}\ \emph {et~al.}(1968)\citenamefont
  {Pilkington}, \citenamefont {Hewish}, \citenamefont {Bell},\ and\
  \citenamefont {Cole}}]{phbc68}%
  \BibitemOpen
  \bibfield  {author} {\bibinfo {author} {\bibfnamefont {J.~D.}\ \bibnamefont
  {Pilkington}}, \bibinfo {author} {\bibfnamefont {A.}~\bibnamefont {Hewish}},
  \bibinfo {author} {\bibfnamefont {S.~J.}\ \bibnamefont {Bell}}, and\ \bibinfo
  {author} {\bibfnamefont {T.~W.}\ \bibnamefont {Cole}},\ }\bibfield  {title}
  {\enquote {\bibinfo {title} {Observations of some further pulsed radio
  sources},}\ }\href {https://doi.org/10.1038/218126a0} {\bibfield  {journal}
  {\bibinfo  {journal} {Nature}\ }\textbf {\bibinfo {volume} {218}},\ \bibinfo
  {pages} {126--+} (\bibinfo {year} {1968})}\BibitemShut {NoStop}%
\bibitem [{\citenamefont {Tauris}\ \emph {et~al.}(1994)\citenamefont {Tauris},
  \citenamefont {Nicastro}, \citenamefont {Johnston}, \citenamefont
  {Manchester}, \citenamefont {Bailes}, \citenamefont {Lyne}, \citenamefont
  {Glowacki}, \citenamefont {Lorimer},\ and\ \citenamefont {Damico}}]{tnj+94}%
  \BibitemOpen
  \bibfield  {author} {\bibinfo {author} {\bibfnamefont {T.~M.}\ \bibnamefont
  {Tauris}}, \bibinfo {author} {\bibfnamefont {L.}~\bibnamefont {Nicastro}},
  \bibinfo {author} {\bibfnamefont {S.}~\bibnamefont {Johnston}}, \bibinfo
  {author} {\bibfnamefont {R.~N.}\ \bibnamefont {Manchester}}, \bibinfo
  {author} {\bibfnamefont {M.}~\bibnamefont {Bailes}}, \bibinfo {author}
  {\bibfnamefont {A.~G.}\ \bibnamefont {Lyne}}, \bibinfo {author}
  {\bibfnamefont {J.}~\bibnamefont {Glowacki}}, \bibinfo {author}
  {\bibfnamefont {D.~R.}\ \bibnamefont {Lorimer}}, and\ \bibinfo {author}
  {\bibfnamefont {N.}~\bibnamefont {Damico}},\ }\bibfield  {title} {\enquote
  {\bibinfo {title} {Discovery of psr j0108-1431 - the closest known
  neutron-star},}\ }\href {https://doi.org/10.1086/187391} {\bibfield
  {journal} {\bibinfo  {journal} {Astrophysical Journal}\ }\textbf {\bibinfo
  {volume} {428}},\ \bibinfo {pages} {L53--L55} (\bibinfo {year}
  {1994})}\BibitemShut {NoStop}%
\bibitem [{\citenamefont {Arzoumanian}\ \emph {et~al.}(1994)\citenamefont
  {Arzoumanian}, \citenamefont {Nice}, \citenamefont {Taylor},\ and\
  \citenamefont {Thorsett}}]{antt94}%
  \BibitemOpen
  \bibfield  {author} {\bibinfo {author} {\bibfnamefont {Z.}~\bibnamefont
  {Arzoumanian}}, \bibinfo {author} {\bibfnamefont {D.~J.}\ \bibnamefont
  {Nice}}, \bibinfo {author} {\bibfnamefont {J.~H.}\ \bibnamefont {Taylor}},
  and\ \bibinfo {author} {\bibfnamefont {S.~E.}\ \bibnamefont {Thorsett}},\
  }\bibfield  {title} {\enquote {\bibinfo {title} {Timing behavior of 96 radio
  pulsars},}\ }\href {https://doi.org/10.1086/173760} {\bibfield  {journal}
  {\bibinfo  {journal} {Astrophysical Journal}\ }\textbf {\bibinfo {volume}
  {422}},\ \bibinfo {pages} {671--680} (\bibinfo {year} {1994})}\BibitemShut
  {NoStop}%
\bibitem [{\citenamefont {Large}, \citenamefont {Vaughan},\ and\ \citenamefont
  {Wielebinski}(1969)}]{lvw69a}%
  \BibitemOpen
  \bibfield  {author} {\bibinfo {author} {\bibfnamefont {M.~I.}\ \bibnamefont
  {Large}}, \bibinfo {author} {\bibfnamefont {A.~E.}\ \bibnamefont {Vaughan}},
  and\ \bibinfo {author} {\bibfnamefont {R.}~\bibnamefont {Wielebinski}},\
  }\bibfield  {title} {\enquote {\bibinfo {title} {Highly dispersed pulsar and
  3 others},}\ }\href {https://doi.org/10.1038/2231249a0} {\bibfield  {journal}
  {\bibinfo  {journal} {Nature}\ }\textbf {\bibinfo {volume} {223}},\ \bibinfo
  {pages} {1249--+} (\bibinfo {year} {1969})}\BibitemShut {NoStop}%
\bibitem [{\citenamefont {Craft}, \citenamefont {Lovelace},\ and\ \citenamefont
  {Sutton}(1968)}]{cls68}%
  \BibitemOpen
  \bibfield  {author} {\bibinfo {author} {\bibfnamefont {H.}~\bibnamefont
  {Craft}}, \bibinfo {author} {\bibfnamefont {R.}~\bibnamefont {Lovelace}},
  and\ \bibinfo {author} {\bibfnamefont {J.}~\bibnamefont {Sutton}},\
  }\bibfield  {title} {\enquote {\bibinfo {title} {New pulsar. iau circ.,
  2100},}\ }\href {http://adsabs.harvard.edu/abs/1968IAUC.2100....1C} {\
  \textbf {\bibinfo {volume} {2100}} (\bibinfo {year} {1968})}\BibitemShut
  {NoStop}%
\bibitem [{\citenamefont {Jacoby}\ \emph {et~al.}(2009)\citenamefont {Jacoby},
  \citenamefont {Bailes}, \citenamefont {Ord}, \citenamefont {Edwards},\ and\
  \citenamefont {Kulkarni}}]{jbo+09}%
  \BibitemOpen
  \bibfield  {author} {\bibinfo {author} {\bibfnamefont {B.~A.}\ \bibnamefont
  {Jacoby}}, \bibinfo {author} {\bibfnamefont {M.}~\bibnamefont {Bailes}},
  \bibinfo {author} {\bibfnamefont {S.~M.}\ \bibnamefont {Ord}}, \bibinfo
  {author} {\bibfnamefont {R.~T.}\ \bibnamefont {Edwards}}, and\ \bibinfo
  {author} {\bibfnamefont {S.~R.}\ \bibnamefont {Kulkarni}},\ }\bibfield
  {title} {\enquote {\bibinfo {title} {A large-area survey for radio pulsars at
  high {Galactic} latitudes},}\ }\href
  {https://doi.org/10.1088/0004-637x/699/2/2009} {\bibfield  {journal}
  {\bibinfo  {journal} {Astrophysical Journal}\ }\textbf {\bibinfo {volume}
  {699}},\ \bibinfo {pages} {2009--2016} (\bibinfo {year} {2009})}\BibitemShut
  {NoStop}%
\bibitem [{\citenamefont {Burgay}\ \emph {et~al.}(2006)\citenamefont {Burgay},
  \citenamefont {Joshi}, \citenamefont {D'Amico}, \citenamefont {Possenti},
  \citenamefont {Lyne}, \citenamefont {Manchester}, \citenamefont {McLaughlin},
  \citenamefont {Kramer}, \citenamefont {Camilo},\ and\ \citenamefont
  {Freire}}]{bjd+06}%
  \BibitemOpen
  \bibfield  {author} {\bibinfo {author} {\bibfnamefont {M.}~\bibnamefont
  {Burgay}}, \bibinfo {author} {\bibfnamefont {B.~C.}\ \bibnamefont {Joshi}},
  \bibinfo {author} {\bibfnamefont {N.}~\bibnamefont {D'Amico}}, \bibinfo
  {author} {\bibfnamefont {A.}~\bibnamefont {Possenti}}, \bibinfo {author}
  {\bibfnamefont {A.~G.}\ \bibnamefont {Lyne}}, \bibinfo {author}
  {\bibfnamefont {R.~N.}\ \bibnamefont {Manchester}}, \bibinfo {author}
  {\bibfnamefont {M.~A.}\ \bibnamefont {McLaughlin}}, \bibinfo {author}
  {\bibfnamefont {M.}~\bibnamefont {Kramer}}, \bibinfo {author} {\bibfnamefont
  {F.}~\bibnamefont {Camilo}}, and\ \bibinfo {author} {\bibfnamefont
  {P.~C.~C.}\ \bibnamefont {Freire}},\ }\bibfield  {title} {\enquote {\bibinfo
  {title} {The {Parkes} high-latitude pulsar survey},}\ }\href
  {https://doi.org/10.1111/j.1365-2966.2006.10100.x} {\bibfield  {journal}
  {\bibinfo  {journal} {Monthly Notices of the Royal Astronomical Society}\
  }\textbf {\bibinfo {volume} {368}},\ \bibinfo {pages} {283--292} (\bibinfo
  {year} {2006})}\BibitemShut {NoStop}%
\bibitem [{\citenamefont {Vaughan}(1969)}]{vlw69}%
  \BibitemOpen
  \bibfield  {author} {\bibinfo {author} {\bibfnamefont {L.~M. I. . W.~R.}\
  \bibnamefont {Vaughan}, \bibfnamefont {A.~E.}},\ }\bibfield  {title}
  {\enquote {\bibinfo {title} {Three new pulsars.}}\ }\href
  {https://doi.org/10.1038/222963a0} {\bibfield  {journal} {\bibinfo  {journal}
  {Nature}\ }\textbf {\bibinfo {volume} {222}},\ \bibinfo {pages} {963}
  (\bibinfo {year} {1969})}\BibitemShut {NoStop}%
\bibitem [{\citenamefont {Cole}\ and\ \citenamefont {Pilkington}(1968)}]{cp68}%
  \BibitemOpen
  \bibfield  {author} {\bibinfo {author} {\bibfnamefont {T.~W.}\ \bibnamefont
  {Cole}}and\ \bibinfo {author} {\bibfnamefont {J.~D.}\ \bibnamefont
  {Pilkington}},\ }\bibfield  {title} {\enquote {\bibinfo {title} {Search for
  pulsating radio sources in declination range + 44 degrees delta + 90
  degrees},}\ }\href {https://doi.org/10.1038/219574a0} {\bibfield  {journal}
  {\bibinfo  {journal} {Nature}\ }\textbf {\bibinfo {volume} {219}},\ \bibinfo
  {pages} {574--+} (\bibinfo {year} {1968})}\BibitemShut {NoStop}%
\bibitem [{\citenamefont {Camilo}, \citenamefont {Nice},\ and\ \citenamefont
  {Taylor}(1996)}]{cnt96}%
  \BibitemOpen
  \bibfield  {author} {\bibinfo {author} {\bibfnamefont {F.}~\bibnamefont
  {Camilo}}, \bibinfo {author} {\bibfnamefont {D.~J.}\ \bibnamefont {Nice}},
  and\ \bibinfo {author} {\bibfnamefont {J.~H.}\ \bibnamefont {Taylor}},\
  }\bibfield  {title} {\enquote {\bibinfo {title} {A search for millisecond
  pulsars at {Galactic} latitudes -50 degrees $\textless$ b $\textless$ -20
  degrees},}\ }\href {https://doi.org/10.1086/177103} {\bibfield  {journal}
  {\bibinfo  {journal} {Astrophysical Journal}\ }\textbf {\bibinfo {volume}
  {461}},\ \bibinfo {pages} {812--819} (\bibinfo {year} {1996})}\BibitemShut
  {NoStop}%
\bibitem [{\citenamefont {Camilo}\ and\ \citenamefont {Nice}(1995)}]{cn95}%
  \BibitemOpen
  \bibfield  {author} {\bibinfo {author} {\bibfnamefont {F.}~\bibnamefont
  {Camilo}}and\ \bibinfo {author} {\bibfnamefont {D.~J.}\ \bibnamefont
  {Nice}},\ }\bibfield  {title} {\enquote {\bibinfo {title} {Timing parameters
  of 29 pulsars},}\ }\href {https://doi.org/10.1086/175737} {\bibfield
  {journal} {\bibinfo  {journal} {Astrophysical Journal}\ }\textbf {\bibinfo
  {volume} {445}},\ \bibinfo {pages} {756--761} (\bibinfo {year}
  {1995})}\BibitemShut {NoStop}%
\bibitem [{Note3()}]{Note3}%
  \BibitemOpen
  \bibinfo {note} {\protect \leavevmode {\protect \color {black}We have
  repeated the same analysis using the PSF3 event class which is the best
  quartile direction reconstruction. This does not change the determined
  \protect \textit {m\protect \textsubscript {a}} significantly considering all
  17 PSRs. We therefore retain the FRONT analysis to allow direct comparison
  with [\protect \rev@citealpnum {RN131}].}}\BibitemShut {Stop}%
\bibitem [{Note4()}]{Note4}%
  \BibitemOpen
  \bibinfo {note} {\protect \textit {Fermipy} change log version
  0.12.0}\BibitemShut {NoStop}%
\bibitem [{\citenamefont {{Wood}}\ \emph {et~al.}(2017)\citenamefont {{Wood}},
  \citenamefont {{Caputo}}, \citenamefont {{Charles}}, \citenamefont {{Di
  Mauro}}, \citenamefont {{Magill}},\ and\ \citenamefont {{Jeremy Perkins for
  the Fermi-LAT Collaboration}}}]{2017arXiv170709551W}%
  \BibitemOpen
  \bibfield  {author} {\bibinfo {author} {\bibfnamefont {M.}~\bibnamefont
  {{Wood}}}, \bibinfo {author} {\bibfnamefont {R.}~\bibnamefont {{Caputo}}},
  \bibinfo {author} {\bibfnamefont {E.}~\bibnamefont {{Charles}}}, \bibinfo
  {author} {\bibfnamefont {M.}~\bibnamefont {{Di Mauro}}}, \bibinfo {author}
  {\bibfnamefont {J.}~\bibnamefont {{Magill}}}, and\ \bibinfo {author}
  {\bibnamefont {{Jeremy Perkins for the Fermi-LAT Collaboration}}},\
  }\bibfield  {title} {\enquote {\bibinfo {title} {{Fermipy: An open-source
  Python package for analysis of Fermi-LAT Data}},}\ }\href@noop {} {\bibfield
  {journal} {\bibinfo  {journal} {ArXiv e-prints}\ } (\bibinfo {year}
  {2017})},\ \Eprint {https://arxiv.org/abs/1707.09551} {arXiv:1707.09551
  [astro-ph.IM]} \BibitemShut {NoStop}%
\bibitem [{Note5()}]{Note5}%
  \BibitemOpen
  \bibinfo {note} {As described in the Fermi Science Support Centre link
  https\protect \tmspace +\medmuskip
  {.2222em}//fermi.gsfc.nasa.gov/ssc/data/analysis/scitools/source\protect
  \_models.html}\BibitemShut {NoStop}%
\bibitem [{\citenamefont {Leinson}(2014)}]{RN342}%
  \BibitemOpen
  \bibfield  {author} {\bibinfo {author} {\bibfnamefont {L.~B.}\ \bibnamefont
  {Leinson}},\ }\bibfield  {title} {\enquote {\bibinfo {title} {Axion mass
  limit from observations of the neutron star in cassiopeia a},}\ }\href
  {https://doi.org/10.1088/1475-7516/2014/08/031} {\bibfield  {journal}
  {\bibinfo  {journal} {Journal of Cosmology and Astroparticle Physics}\ ,\
  \bibinfo {pages} {11}} (\bibinfo {year} {2014})}\BibitemShut {NoStop}%
\bibitem [{\citenamefont {Ruster}\ \emph {et~al.}(2005)\citenamefont {Ruster},
  \citenamefont {Werth}, \citenamefont {Buballa}, \citenamefont {Shovkovy},\
  and\ \citenamefont {Rischke}}]{RN261}%
  \BibitemOpen
  \bibfield  {author} {\bibinfo {author} {\bibfnamefont {S.~B.}\ \bibnamefont
  {Ruster}}, \bibinfo {author} {\bibfnamefont {V.}~\bibnamefont {Werth}},
  \bibinfo {author} {\bibfnamefont {M.}~\bibnamefont {Buballa}}, \bibinfo
  {author} {\bibfnamefont {I.~A.}\ \bibnamefont {Shovkovy}}, and\ \bibinfo
  {author} {\bibfnamefont {D.~H.}\ \bibnamefont {Rischke}},\ }\bibfield
  {title} {\enquote {\bibinfo {title} {Phase diagram of neutral quark matter:
  Self-consistent treatment of quark masses},}\ }\href
  {https://doi.org/10.1103/PhysRevD.72.034004} {\bibfield  {journal} {\bibinfo
  {journal} {Physical Review D}\ }\textbf {\bibinfo {volume} {72}},\ \bibinfo
  {pages} {13} (\bibinfo {year} {2005})}\BibitemShut {NoStop}%
\bibitem [{\citenamefont {Shen}\ \emph {et~al.}(1998)\citenamefont {Shen},
  \citenamefont {Toki}, \citenamefont {Oyamatsu},\ and\ \citenamefont
  {Sumiyoshi}}]{RN291}%
  \BibitemOpen
  \bibfield  {author} {\bibinfo {author} {\bibfnamefont {H.}~\bibnamefont
  {Shen}}, \bibinfo {author} {\bibfnamefont {H.}~\bibnamefont {Toki}}, \bibinfo
  {author} {\bibfnamefont {K.}~\bibnamefont {Oyamatsu}}, and\ \bibinfo {author}
  {\bibfnamefont {K.}~\bibnamefont {Sumiyoshi}},\ }\bibfield  {title} {\enquote
  {\bibinfo {title} {Relativistic equation of state of nuclear matter for
  supernova and neutron star},}\ }\href
  {https://doi.org/10.1016/s0375-9474(98)00236-x} {\bibfield  {journal}
  {\bibinfo  {journal} {Nuclear Physics A}\ }\textbf {\bibinfo {volume}
  {637}},\ \bibinfo {pages} {435--450} (\bibinfo {year} {1998})}\BibitemShut
  {NoStop}%
\bibitem [{\citenamefont {Akmal}, \citenamefont {Pandharipande},\ and\
  \citenamefont {Ravenhall}(1998)}]{RN292}%
  \BibitemOpen
  \bibfield  {author} {\bibinfo {author} {\bibfnamefont {A.}~\bibnamefont
  {Akmal}}, \bibinfo {author} {\bibfnamefont {V.~R.}\ \bibnamefont
  {Pandharipande}}, and\ \bibinfo {author} {\bibfnamefont {D.~G.}\ \bibnamefont
  {Ravenhall}},\ }\bibfield  {title} {\enquote {\bibinfo {title} {Equation of
  state of nucleon matter and neutron star structure},}\ }\href
  {https://doi.org/10.1103/PhysRevC.58.1804} {\bibfield  {journal} {\bibinfo
  {journal} {Physical Review C}\ }\textbf {\bibinfo {volume} {58}},\ \bibinfo
  {pages} {1804--1828} (\bibinfo {year} {1998})}\BibitemShut {NoStop}%
\bibitem [{\citenamefont {Negreiros}, \citenamefont {Dexheimer},\ and\
  \citenamefont {Schramm}(2012)}]{RN262}%
  \BibitemOpen
  \bibfield  {author} {\bibinfo {author} {\bibfnamefont {R.}~\bibnamefont
  {Negreiros}}, \bibinfo {author} {\bibfnamefont {V.~A.}\ \bibnamefont
  {Dexheimer}}, and\ \bibinfo {author} {\bibfnamefont {S.}~\bibnamefont
  {Schramm}},\ }\bibfield  {title} {\enquote {\bibinfo {title} {Quark core
  impact on hybrid star cooling},}\ }\href
  {https://doi.org/10.1103/PhysRevC.85.035805} {\bibfield  {journal} {\bibinfo
  {journal} {Physical Review C}\ }\textbf {\bibinfo {volume} {85}},\ \bibinfo
  {pages} {7} (\bibinfo {year} {2012})}\BibitemShut {NoStop}%
\bibitem [{\citenamefont {Larson}\ and\ \citenamefont {Link}(1999)}]{RN294}%
  \BibitemOpen
  \bibfield  {author} {\bibinfo {author} {\bibfnamefont {M.~B.}\ \bibnamefont
  {Larson}}and\ \bibinfo {author} {\bibfnamefont {B.}~\bibnamefont {Link}},\
  }\bibfield  {title} {\enquote {\bibinfo {title} {Superfluid friction and
  late-time thermal evolution of neutron stars},}\ }\href
  {https://doi.org/10.1086/307532} {\bibfield  {journal} {\bibinfo  {journal}
  {Astrophysical Journal}\ }\textbf {\bibinfo {volume} {521}},\ \bibinfo
  {pages} {271--280} (\bibinfo {year} {1999})}\BibitemShut {NoStop}%
\bibitem [{\citenamefont {Pavlov}, \citenamefont {Stringfellow},\ and\
  \citenamefont {Cordova}(1996)}]{RN295}%
  \BibitemOpen
  \bibfield  {author} {\bibinfo {author} {\bibfnamefont {G.~G.}\ \bibnamefont
  {Pavlov}}, \bibinfo {author} {\bibfnamefont {G.~S.}\ \bibnamefont
  {Stringfellow}}, and\ \bibinfo {author} {\bibfnamefont {F.~A.}\ \bibnamefont
  {Cordova}},\ }\bibfield  {title} {\enquote {\bibinfo {title} {Hubble space
  telescope observations of isolated pulsars},}\ }\href
  {https://doi.org/10.1086/177612} {\bibfield  {journal} {\bibinfo  {journal}
  {Astrophysical Journal}\ }\textbf {\bibinfo {volume} {467}},\ \bibinfo
  {pages} {370--+} (\bibinfo {year} {1996})}\BibitemShut {NoStop}%
\bibitem [{\citenamefont {Pavlov}\ \emph {et~al.}(2017)\citenamefont {Pavlov},
  \citenamefont {Rangelov}, \citenamefont {Kargaltsev}, \citenamefont
  {Reisenegger}, \citenamefont {Guillot},\ and\ \citenamefont {Reyes}}]{RN256}%
  \BibitemOpen
  \bibfield  {author} {\bibinfo {author} {\bibfnamefont {G.~G.}\ \bibnamefont
  {Pavlov}}, \bibinfo {author} {\bibfnamefont {B.}~\bibnamefont {Rangelov}},
  \bibinfo {author} {\bibfnamefont {O.}~\bibnamefont {Kargaltsev}}, \bibinfo
  {author} {\bibfnamefont {A.}~\bibnamefont {Reisenegger}}, \bibinfo {author}
  {\bibfnamefont {S.}~\bibnamefont {Guillot}}, and\ \bibinfo {author}
  {\bibfnamefont {C.}~\bibnamefont {Reyes}},\ }\bibfield  {title} {\enquote
  {\bibinfo {title} {Old but still warm: Far-uv detection of psr b0950+08},}\
  }\href {https://doi.org/10.3847/1538-4357/aa947c} {\bibfield  {journal}
  {\bibinfo  {journal} {Astrophysical Journal}\ }\textbf {\bibinfo {volume}
  {850}},\ \bibinfo {pages} {7} (\bibinfo {year} {2017})}\BibitemShut {NoStop}%
\bibitem [{\citenamefont {Gudmundsson}, \citenamefont {Pethick},\ and\
  \citenamefont {Epstein}(1982)}]{RN301}%
  \BibitemOpen
  \bibfield  {author} {\bibinfo {author} {\bibfnamefont {E.~H.}\ \bibnamefont
  {Gudmundsson}}, \bibinfo {author} {\bibfnamefont {C.~J.}\ \bibnamefont
  {Pethick}}, and\ \bibinfo {author} {\bibfnamefont {R.~I.}\ \bibnamefont
  {Epstein}},\ }\bibfield  {title} {\enquote {\bibinfo {title} {Neutron star
  envelopes},}\ }\href {https://doi.org/10.1086/183840} {\bibfield  {journal}
  {\bibinfo  {journal} {Astrophysical Journal}\ }\textbf {\bibinfo {volume}
  {259}},\ \bibinfo {pages} {L19--L23} (\bibinfo {year} {1982})}\BibitemShut
  {NoStop}%
\bibitem [{\citenamefont {Nomoto}\ and\ \citenamefont {Tsuruta}(1987)}]{RN344}%
  \BibitemOpen
  \bibfield  {author} {\bibinfo {author} {\bibfnamefont {K.}~\bibnamefont
  {Nomoto}}and\ \bibinfo {author} {\bibfnamefont {S.}~\bibnamefont {Tsuruta}},\
  }\bibfield  {title} {\enquote {\bibinfo {title} {Cooling of neutron-stars -
  effects of the finite-time scale of thermal conduction},}\ }\href
  {https://doi.org/10.1086/164914} {\bibfield  {journal} {\bibinfo  {journal}
  {Astrophysical Journal}\ }\textbf {\bibinfo {volume} {312}},\ \bibinfo
  {pages} {711--726} (\bibinfo {year} {1987})}\BibitemShut {NoStop}%
\bibitem [{\citenamefont {Yakovlev}\ and\ \citenamefont
  {Pethick}(2004)}]{PSR_REVIEW_YAKOVLEV}%
  \BibitemOpen
  \bibfield  {author} {\bibinfo {author} {\bibfnamefont {D.}~\bibnamefont
  {Yakovlev}}and\ \bibinfo {author} {\bibfnamefont {C.}~\bibnamefont
  {Pethick}},\ }\bibfield  {title} {\enquote {\bibinfo {title} {Neutron star
  cooling},}\ }\href@noop {} {\bibfield  {journal} {\bibinfo  {journal} {Annual
  Review of Astronomy and Astrophysics}\ }\textbf {\bibinfo {volume} {42}},\
  \bibinfo {pages} {169--210} (\bibinfo {year} {2004})}\BibitemShut {NoStop}%
\bibitem [{\citenamefont {Heinke}\ and\ \citenamefont {Ho}(2010)}]{RN343}%
  \BibitemOpen
  \bibfield  {author} {\bibinfo {author} {\bibfnamefont {C.~O.}\ \bibnamefont
  {Heinke}}and\ \bibinfo {author} {\bibfnamefont {W.~C.~G.}\ \bibnamefont
  {Ho}},\ }\bibfield  {title} {\enquote {\bibinfo {title} {Direct observation
  of the cooling of the cassiopeia a neutron star},}\ }\href
  {https://doi.org/10.1088/2041-8205/719/2/l167} {\bibfield  {journal}
  {\bibinfo  {journal} {Astrophysical Journal Letters}\ }\textbf {\bibinfo
  {volume} {719}},\ \bibinfo {pages} {L167--L171} (\bibinfo {year}
  {2010})}\BibitemShut {NoStop}%
\bibitem [{\citenamefont {Sumiyoshi}\ \emph {et~al.}(2005)\citenamefont
  {Sumiyoshi}, \citenamefont {Yamada}, \citenamefont {Suzuki}, \citenamefont
  {Shen}, \citenamefont {Chiba},\ and\ \citenamefont {Toki}}]{RN297}%
  \BibitemOpen
  \bibfield  {author} {\bibinfo {author} {\bibfnamefont {K.}~\bibnamefont
  {Sumiyoshi}}, \bibinfo {author} {\bibfnamefont {S.}~\bibnamefont {Yamada}},
  \bibinfo {author} {\bibfnamefont {H.}~\bibnamefont {Suzuki}}, \bibinfo
  {author} {\bibfnamefont {H.}~\bibnamefont {Shen}}, \bibinfo {author}
  {\bibfnamefont {S.}~\bibnamefont {Chiba}}, and\ \bibinfo {author}
  {\bibfnamefont {H.}~\bibnamefont {Toki}},\ }\bibfield  {title} {\enquote
  {\bibinfo {title} {Postbounce evolution of core-collapse supernovae:
  Long-term effects of the equation of state},}\ }\href
  {https://doi.org/10.1086/431788} {\bibfield  {journal} {\bibinfo  {journal}
  {Astrophysical Journal}\ }\textbf {\bibinfo {volume} {629}},\ \bibinfo
  {pages} {922--932} (\bibinfo {year} {2005})}\BibitemShut {NoStop}%
\bibitem [{\citenamefont {Pons}\ \emph {et~al.}(1999)\citenamefont {Pons},
  \citenamefont {Reddy}, \citenamefont {Prakash}, \citenamefont {Lattimer},\
  and\ \citenamefont {Miralles}}]{RN296}%
  \BibitemOpen
  \bibfield  {author} {\bibinfo {author} {\bibfnamefont {J.~A.}\ \bibnamefont
  {Pons}}, \bibinfo {author} {\bibfnamefont {S.}~\bibnamefont {Reddy}},
  \bibinfo {author} {\bibfnamefont {M.}~\bibnamefont {Prakash}}, \bibinfo
  {author} {\bibfnamefont {J.~M.}\ \bibnamefont {Lattimer}}, and\ \bibinfo
  {author} {\bibfnamefont {J.~A.}\ \bibnamefont {Miralles}},\ }\bibfield
  {title} {\enquote {\bibinfo {title} {Evolution of proto-neutron stars},}\
  }\href {https://doi.org/10.1086/306889} {\bibfield  {journal} {\bibinfo
  {journal} {Astrophysical Journal}\ }\textbf {\bibinfo {volume} {513}},\
  \bibinfo {pages} {780--804} (\bibinfo {year} {1999})}\BibitemShut {NoStop}%
\bibitem [{\citenamefont {Nakazato}, \citenamefont {Suzuki},\ and\
  \citenamefont {Togashi}(2018)}]{Nakazato2018}%
  \BibitemOpen
  \bibfield  {author} {\bibinfo {author} {\bibfnamefont {K.}~\bibnamefont
  {Nakazato}}, \bibinfo {author} {\bibfnamefont {H.}~\bibnamefont {Suzuki}},
  and\ \bibinfo {author} {\bibfnamefont {H.}~\bibnamefont {Togashi}},\
  }\bibfield  {title} {\enquote {\bibinfo {title} {Heavy nuclei as thermal
  insulation for protoneutron stars},}\ }\href
  {https://doi.org/10.1103/PhysRevC.97.035804} {\bibfield  {journal} {\bibinfo
  {journal} {Phys. Rev. C}\ }\textbf {\bibinfo {volume} {97}},\ \bibinfo
  {pages} {035804} (\bibinfo {year} {2018})}\BibitemShut {NoStop}%
\bibitem [{\citenamefont {Zhu}, \citenamefont {Lu},\ and\ \citenamefont
  {Wang}(2018)}]{RN265}%
  \BibitemOpen
  \bibfield  {author} {\bibinfo {author} {\bibfnamefont {L.~G.}\ \bibnamefont
  {Zhu}}, \bibinfo {author} {\bibfnamefont {J.~L.}\ \bibnamefont {Lu}}, and\
  \bibinfo {author} {\bibfnamefont {L.}~\bibnamefont {Wang}},\ }\bibfield
  {title} {\enquote {\bibinfo {title} {Effects of temperature on the structure
  of neutron stars at high temperature},}\ }\href
  {https://doi.org/10.1007/s10714-017-2327-3} {\bibfield  {journal} {\bibinfo
  {journal} {General Relativity and Gravitation}\ }\textbf {\bibinfo {volume}
  {50}},\ \bibinfo {pages} {18} (\bibinfo {year} {2018})}\BibitemShut {NoStop}%
\bibitem [{\citenamefont {Perna}\ \emph {et~al.}(2012)\citenamefont {Perna},
  \citenamefont {Ho}, \citenamefont {Verde}, \citenamefont {van Adelsberg},\
  and\ \citenamefont {Jimenez}}]{RN227}%
  \BibitemOpen
  \bibfield  {author} {\bibinfo {author} {\bibfnamefont {R.}~\bibnamefont
  {Perna}}, \bibinfo {author} {\bibfnamefont {W.~C.~G.}\ \bibnamefont {Ho}},
  \bibinfo {author} {\bibfnamefont {L.}~\bibnamefont {Verde}}, \bibinfo
  {author} {\bibfnamefont {M.}~\bibnamefont {van Adelsberg}}, and\ \bibinfo
  {author} {\bibfnamefont {R.}~\bibnamefont {Jimenez}},\ }\bibfield  {title}
  {\enquote {\bibinfo {title} {Signatures of photon-axion conversion in the
  thermal spectra and polarization of neutron stars},}\ }\href
  {https://doi.org/10.1088/0004-637x/748/2/116} {\bibfield  {journal} {\bibinfo
   {journal} {Astrophysical Journal}\ }\textbf {\bibinfo {volume} {748}},\
  \bibinfo {pages} {17} (\bibinfo {year} {2012})}\BibitemShut {NoStop}%
\bibitem [{\citenamefont {Fortin}\ and\ \citenamefont {Sinha}(2018)}]{RN299}%
  \BibitemOpen
  \bibfield  {author} {\bibinfo {author} {\bibfnamefont {J.~F.}\ \bibnamefont
  {Fortin}}and\ \bibinfo {author} {\bibfnamefont {K.}~\bibnamefont {Sinha}},\
  }\bibfield  {title} {\enquote {\bibinfo {title} {Constraining
  axion-like-particles with hard x-ray emission from magnetars},}\ }\href
  {https://doi.org/10.1007/jhep06(2018)048} {\bibfield  {journal} {\bibinfo
  {journal} {Journal of High Energy Physics}\ ,\ \bibinfo {pages} {22}}
  (\bibinfo {year} {2018})}\BibitemShut {NoStop}%
\bibitem [{\citenamefont {Rando}(2017)}]{AMEGO}%
  \BibitemOpen
  \bibfield  {author} {\bibinfo {author} {\bibfnamefont {R.}~\bibnamefont
  {Rando}},\ }\bibfield  {title} {\enquote {\bibinfo {title} {The all-sky
  medium energy gamma-ray observatory},}\ }\href
  {http://stacks.iop.org/1748-0221/12/i=11/a=C11024} {\bibfield  {journal}
  {\bibinfo  {journal} {Journal of Instrumentation}\ }\textbf {\bibinfo
  {volume} {12}},\ \bibinfo {pages} {C11024} (\bibinfo {year}
  {2017})}\BibitemShut {NoStop}%
\bibitem [{\citenamefont {Angelis}\ \emph {et~al.}(2018)\citenamefont
  {Angelis}, \citenamefont {Tatischeff}, \citenamefont {Grenier}, \citenamefont
  {McEnery}, \citenamefont {Mallamaci}, \citenamefont {Tavani}, \citenamefont
  {Oberlack}, \citenamefont {Hanlon}, \citenamefont {Walter}, \citenamefont
  {Argan}, \citenamefont {Ballmoos}, \citenamefont {Bulgarelli}, \citenamefont
  {Bykov}, \citenamefont {Hernanz}, \citenamefont {Kanbach}, \citenamefont
  {Kuvvetli}, \citenamefont {Pearce}, \citenamefont {Zdziarski}, \citenamefont
  {Conrad}, \citenamefont {Ghisellini}, \citenamefont {Harding}, \citenamefont
  {Isern}, \citenamefont {Leising}, \citenamefont {Longo}, \citenamefont
  {Madejski}, \citenamefont {Martinez}, \citenamefont {Mazziotta},
  \citenamefont {Paredes}, \citenamefont {Pohl}, \citenamefont {Rando},
  \citenamefont {Razzano}, \citenamefont {Aboudan}, \citenamefont {Ackermann},
  \citenamefont {Addazi}, \citenamefont {Ajello}, \citenamefont {Albertus}
  \emph {et~al.}}]{DEANGELIS20181}%
  \BibitemOpen
  \bibfield  {author} {\bibinfo {author} {\bibfnamefont {A.~D.}\ \bibnamefont
  {Angelis}}, \bibinfo {author} {\bibfnamefont {V.}~\bibnamefont {Tatischeff}},
  \bibinfo {author} {\bibfnamefont {I.}~\bibnamefont {Grenier}}, \bibinfo
  {author} {\bibfnamefont {J.}~\bibnamefont {McEnery}}, \bibinfo {author}
  {\bibfnamefont {M.}~\bibnamefont {Mallamaci}}, \bibinfo {author}
  {\bibfnamefont {M.}~\bibnamefont {Tavani}}, \bibinfo {author} {\bibfnamefont
  {U.}~\bibnamefont {Oberlack}}, \bibinfo {author} {\bibfnamefont
  {L.}~\bibnamefont {Hanlon}}, \bibinfo {author} {\bibfnamefont
  {R.}~\bibnamefont {Walter}}, \bibinfo {author} {\bibfnamefont
  {A.}~\bibnamefont {Argan}}, \bibinfo {author} {\bibfnamefont {P.~V.}\
  \bibnamefont {Ballmoos}}, \bibinfo {author} {\bibfnamefont {A.}~\bibnamefont
  {Bulgarelli}}, \bibinfo {author} {\bibfnamefont {A.}~\bibnamefont {Bykov}},
  \bibinfo {author} {\bibfnamefont {M.}~\bibnamefont {Hernanz}}, \bibinfo
  {author} {\bibfnamefont {G.}~\bibnamefont {Kanbach}}, \bibinfo {author}
  {\bibfnamefont {I.}~\bibnamefont {Kuvvetli}}, \bibinfo {author}
  {\bibfnamefont {M.}~\bibnamefont {Pearce}}, \bibinfo {author} {\bibfnamefont
  {A.}~\bibnamefont {Zdziarski}}, \bibinfo {author} {\bibfnamefont
  {J.}~\bibnamefont {Conrad}}, \bibinfo {author} {\bibfnamefont
  {G.}~\bibnamefont {Ghisellini}}, \bibinfo {author} {\bibfnamefont
  {A.}~\bibnamefont {Harding}}, \bibinfo {author} {\bibfnamefont
  {J.}~\bibnamefont {Isern}}, \bibinfo {author} {\bibfnamefont
  {M.}~\bibnamefont {Leising}}, \bibinfo {author} {\bibfnamefont
  {F.}~\bibnamefont {Longo}}, \bibinfo {author} {\bibfnamefont
  {G.}~\bibnamefont {Madejski}}, \bibinfo {author} {\bibfnamefont
  {M.}~\bibnamefont {Martinez}}, \bibinfo {author} {\bibfnamefont
  {M.}~\bibnamefont {Mazziotta}}, \bibinfo {author} {\bibfnamefont
  {J.}~\bibnamefont {Paredes}}, \bibinfo {author} {\bibfnamefont
  {M.}~\bibnamefont {Pohl}}, \bibinfo {author} {\bibfnamefont {R.}~\bibnamefont
  {Rando}}, \bibinfo {author} {\bibfnamefont {M.}~\bibnamefont {Razzano}},
  \bibinfo {author} {\bibfnamefont {A.}~\bibnamefont {Aboudan}}, \bibinfo
  {author} {\bibfnamefont {M.}~\bibnamefont {Ackermann}}, \bibinfo {author}
  {\bibfnamefont {A.}~\bibnamefont {Addazi}}, \bibinfo {author} {\bibfnamefont
  {M.}~\bibnamefont {Ajello}}, \bibinfo {author} {\bibfnamefont
  {C.}~\bibnamefont {Albertus}},  \emph {et~al.},\ }\bibfield  {title}
  {\enquote {\bibinfo {title} {Science with e-astrogam: A space mission for
  mev–gev gamma-ray astrophysics},}\ }\href
  {https://doi.org/https://doi.org/10.1016/j.jheap.2018.07.001} {\bibfield
  {journal} {\bibinfo  {journal} {Journal of High Energy Astrophysics}\
  }\textbf {\bibinfo {volume} {19}},\ \bibinfo {pages} {1 -- 106} (\bibinfo
  {year} {2018})}\BibitemShut {NoStop}%
\bibitem [{\citenamefont {{Wenger}}\ \emph {et~al.}(2000)\citenamefont
  {{Wenger}}, \citenamefont {{Ochsenbein}}, \citenamefont {{Egret}},
  \citenamefont {{Dubois}}, \citenamefont {{Bonnarel}}, \citenamefont
  {{Borde}}, \citenamefont {{Genova}}, \citenamefont {{Jasniewicz}},
  \citenamefont {{Lalo{\"e}}}, \citenamefont {{Lesteven}},\ and\ \citenamefont
  {{Monier}}}]{2000A&AS..143....9W}%
  \BibitemOpen
  \bibfield  {author} {\bibinfo {author} {\bibfnamefont {M.}~\bibnamefont
  {{Wenger}}}, \bibinfo {author} {\bibfnamefont {F.}~\bibnamefont
  {{Ochsenbein}}}, \bibinfo {author} {\bibfnamefont {D.}~\bibnamefont
  {{Egret}}}, \bibinfo {author} {\bibfnamefont {P.}~\bibnamefont {{Dubois}}},
  \bibinfo {author} {\bibfnamefont {F.}~\bibnamefont {{Bonnarel}}}, \bibinfo
  {author} {\bibfnamefont {S.}~\bibnamefont {{Borde}}}, \bibinfo {author}
  {\bibfnamefont {F.}~\bibnamefont {{Genova}}}, \bibinfo {author}
  {\bibfnamefont {G.}~\bibnamefont {{Jasniewicz}}}, \bibinfo {author}
  {\bibfnamefont {S.}~\bibnamefont {{Lalo{\"e}}}}, \bibinfo {author}
  {\bibfnamefont {S.}~\bibnamefont {{Lesteven}}}, and\ \bibinfo {author}
  {\bibfnamefont {R.}~\bibnamefont {{Monier}}},\ }\bibfield  {title} {\enquote
  {\bibinfo {title} {{The SIMBAD astronomical database. The CDS reference
  database for astronomical objects}},}\ }\href
  {https://doi.org/10.1051/aas:2000332} {\bibfield  {journal} {\bibinfo
  {journal} {Astronomy and Astrophysics Supplement}\ }\textbf {\bibinfo
  {volume} {143}},\ \bibinfo {pages} {9--22} (\bibinfo {year} {2000})},\
  \Eprint {https://arxiv.org/abs/astro-ph/0002110} {astro-ph/0002110}
  \BibitemShut {NoStop}%
\end{thebibliography}%

\end{document}